\documentclass[sigplan,nonacm]{acmart}

\begin{document}
\title{HBFlex: A Flexible Memory System for Bridging Fine-Grained LLM States and  Coarse-Grained HBF Parallel Execution}

\begin{abstract}
Large language models (LLMs) require increasing memory capacity to
accommodate growing model weights and KV caches. High-Bandwidth Flash
(HBF) offers high memory density and aggregate read bandwidth through
massive plane-level parallelism, making it an attractive option for LLM
serving. However, serving LLMs entirely from HBF
introduces three challenges: fine-grained KV reads create placement and
access imbalance, incremental writes interfere with foreground reads,
and mixed KV lifetimes amplify garbage collection. Hybrid HBM/HBF
designs retain HBM to support dynamic KV management, but this allocation
reduces the HBF resources available under a fixed packaging budget,
limiting aggregate HBF bandwidth.

We present HBFlex, a full-HBF memory system with coordinated
optimizations for KV reads, writes, and reclamation. HBFlex balances KV
placement and attention accesses to improve plane utilization. It
aggregates incremental updates and schedules writeback within
sufficiently long compute windows to reduce write--read interference.
It also combines lifetime-guided block packing with deferred reclamation
to reduce valid-page migration. We evaluate HBFlex through trace-driven
simulation across different configurations. HBFlex achieves average
throughput speedups of up to 1.58$\times$ over FlashAccel and
3.30$\times$ over H3, benefiting from higher HBF bandwidth and more
efficient management of dynamic KV-cache reads, writes, and erases.
\end{abstract}

\author{Shuzhang Zhong}
%\authornote{Co-first authors.}
\authornotemark[1]
\affiliation{%
  \institution{Peking University}
  \city{Beijing}
  \country{China}
}

\author{Weikai Xu}
\authornotemark[1]
\affiliation{%
  \institution{HKUST}
  \city{Hong Kong}
  \country{China}
}

\author{Yifan Zhou}
\affiliation{%
  \institution{Peking University}
  \city{Beijing}
  \country{China}
}

\author{Tongbin Zhao}
\affiliation{%
  \institution{Peking University}
  \city{Beijing}
  \country{China}
}

\author{Tenghao Zhao}
\affiliation{%
  \institution{Peking University}
  \city{Beijing}
  \country{China}
}

\author{Yifei Kang}
\affiliation{%
  \institution{Alibaba Group}
  \city{Hangzhou}
  \country{China}
}

\author{Cunyin Chang}
\affiliation{%
  \institution{Alibaba Group}
  \city{Hangzhou}
  \country{China}
}

\author{Shu Li}
\affiliation{%
  \institution{Alibaba Group}
  \city{Hangzhou}
  \country{China}
}

\author{Guangyu Sun}
\affiliation{%
  \institution{Peking University}
  \city{Beijing}
  \country{China}
}

\author{Meng Li}
\authornotemark[2]
%\authornote{Corresponding author.}
\affiliation{%
  \institution{Peking University}
  \city{Beijing}
  \country{China}
}
\email{meng.li@pku.edu.cn}

\settopmatter{authorsperrow=4}
\maketitle % should come after the abstract
\makeatletter
\fancyhead[LO]{\ACM@linecountL}
\fancyhead[RE]{\ACM@linecountR}
\makeatother

\section{Introduction}
\label{sec:introduction}

Large language models (LLMs) and agentic applications~\cite{park2023generative,yao2023react,wang2023voyager} are placing increasing pressure on accelerator memory capacity.
Model sizes continue to grow, while long contexts~\cite{zhang2023h2o,xiao2024efficient}, multi-turn interactions~\cite{deshpande2025multichallenge,yi2025survey,zheng2023judging}, and concurrent requests increase the amount of key--value (KV) state that serving
systems must retain.
For example, DeepSeek-V4-Pro~\cite{deepseekai2026deepseekv4} contains 1.6 trillion parameters and supports a one-million-token context window.
Preserving useful KV histories also
avoids repeated prefill computation across related requests~\cite{zheng2024sglang,qin2025mooncake}.
However, the limited capacity of high-bandwidth memory (HBM) makes it difficult to accommodate both large models and their growing KV caches on a single accelerator.
Distributing this state across more GPUs relieves capacity pressure but adds communication and coordination overhead,
increasing the cost of serving~\cite{
aminabadi2022deepspeedinference,li2023alpaserve,patel2024splitwise,
zhong2024distserve}.

High-Bandwidth Flash (HBF) offers a promising path to expanding accelerator-local memory capacity by combining the high density of NAND flash with vertically stacked dies and a wide accelerator interface~\cite{maChallengesResearchDirections2026,
sonExploringHighBandwidthFlash2026,kyungHighBandwidthFlashKV2026}.
Although an individual NAND access is slower than a DRAM access, thousands of independent planes can operate concurrently, allowing HBF to target read bandwidth comparable to HBM with substantially greater capacity density.
Its bandwidth advantage therefore depends on exposing sufficient plane-level parallelism and distributing useful work evenly across many planes.

Recent studies have explored hybrid HBM/HBF organizations that capture HBF's capacity benefits while retaining HBM for state that benefits from low latency and efficient updates.
These proposals span \emph{co-located} designs with separate HBM and HBF stacks beside the GPU~\cite{wangFlashAccelLeveragingHighBandwidth2026,
leeMemVillageHybridHBFHBM2025}, \emph{cascaded} designs that connect HBF through the HBM base die~\cite{haH3HybridArchitecture2026,
wangFlashAccelLeveragingHighBandwidth2026,parkHBMHBFCentricMemoryPooling2026}, and \emph{integrated-stack} designs that combine DRAM and flash dies within one stack~\cite{yin2026potentialapplicationshbfllm}.
A common approach places read-mostly state, such as weights and precomputed shared KV, in HBF while retaining dynamic request KV in HBM~\cite{haH3HybridArchitecture2026}.

\begin{figure}[t]
    \centering
    \includegraphics[width=\linewidth]{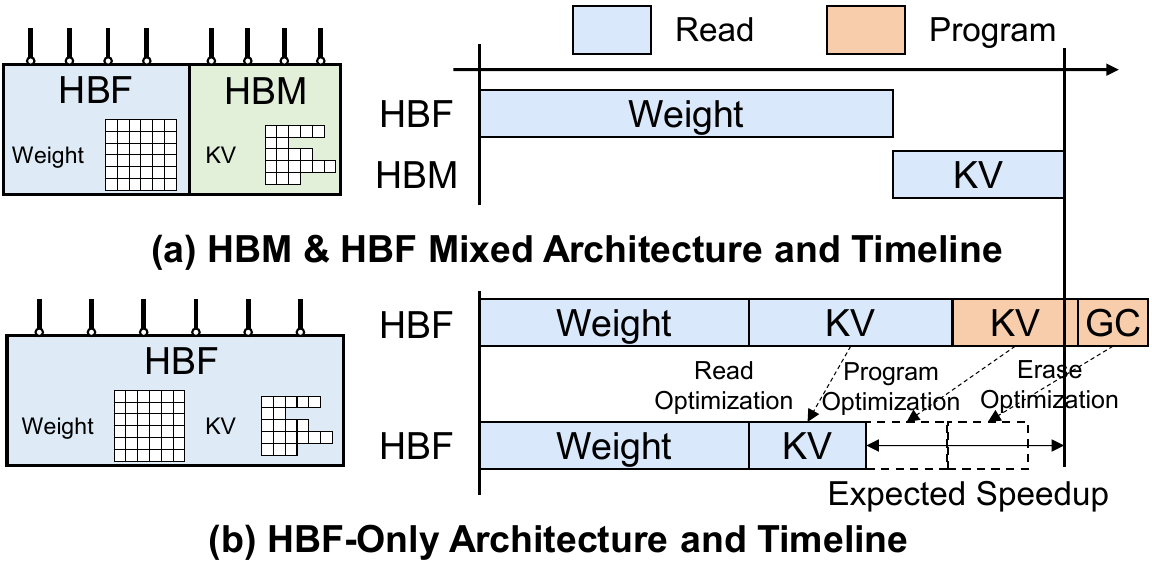}
    \caption{From hybrid HBM/HBF serving to full-HBF serving.
    (a) A hybrid organization separates weight reads and dynamic KV
    accesses across HBF and HBM. (b) A full-HBF organization provides
    more HBF resources, but realizing its performance potential requires
    coordinated optimization of KV reads, programs, and block erases.}
    \Description{Two memory organizations and their execution timelines
    contrast hybrid HBM/HBF serving with full-HBF serving. The latter
    requires read, program, and erase optimizations to realize the
    expected speedup.}
    \label{fig:introduction}
\end{figure}

However, under a fixed package budget, retaining HBM comes at the expense of HBF resources.
HBM occupies stack locations or die resources that could otherwise accommodate additional HBF planes.
Because HBF derives aggregate bandwidth from these planes, allocating fewer resources to HBF limits both its capacity and its available read parallelism.
In particular, the bandwidth available for HBF-resident weights remains constrained even when HBM efficiently serves KV accesses, as illustrated in Figure~\ref{fig:introduction}(a).

Dedicating the entire memory budget to HBF eliminates this resource trade-off but places dynamic KV reads, writes, and reclamation on the same flash device.
Fine-grained accesses can leave much of the nominal bandwidth unused, incremental updates can block subsequent reads, and garbage collection (GC) can introduce additional data movement and consume flash endurance~\cite{ocp_hbf_architecture_v070,dirik2009performance}.
Consequently, simply moving KV into HBF does not ensure that its additional planes translate into better serving performance~\cite{liHBFSucksFullStack2026}.

Exploiting a full-HBF organization therefore requires a memory system that reconciles the logical behavior of KV caches with the physical constraints of flash.
As Figure~\ref{fig:introduction}(b) illustrates,  the system must coordinate read, program, and erase operations to preserve the parallelism gained by dedicating more resources to HBF.
We identify three challenges that prevent a direct mapping of conventional KV management onto this organization.

\textbf{Challenge 1: Fine-grained KV cache creates read imbalance.}
This imbalance arises at two distinct levels.
Request completion and reactivation can create \textbf{placement
imbalance}. They change the active KV working set without necessarily
moving its pages, disrupting a previously balanced placement.
Balancing new allocations therefore does not ensure balance as request
activity changes. Even with balanced placement, fine-grained reads can
create \textbf{access imbalance}. Each attention invocation reads only
a subset of KV pages. PagedAttention~\cite{kwon2023pagedattention} and chunked CTA scheduling~\cite{agrawal2023sarathi, agrawal2024sarathiserve} can
select pages concentrated on a few planes. These collisions require
additional read waves while other planes remain idle.

\textbf{Challenge 2: Incremental KV updates create write--read interference.}
Incremental KV updates require buffering to form writes that exploit
plane parallelism. However, each program occupies its target plane
for approximately 75~$\mu$s and cannot be preempted. Aggregation alone
therefore cannot prevent interference with foreground reads.
Interleaved weight and KV reads divide computation into short gaps
that may not hide a full program. Deferring writes avoids immediate
conflicts, but buffer pressure eventually forces writeback. Efficient
writeback must coordinate both write aggregation and program timing.

\textbf{Challenge 3: Interleaved KV lifetimes create erase amplification.}
Raw KV writes remain within the estimated endurance budget under our
workload assumptions. However, block-level erase requires valid pages
to be migrated when KV pages with different lifetimes share a block.
Our characterization of DeepSeek-V4-Pro shows an average GC write amplification of
approximately 30$\times$, exceeding the available endurance margins.
Isolating requests in dedicated blocks simplifies erase management,
but constrains plane-level striping. KV placement must therefore
balance read parallelism against lifetime locality.

We present \emph{HBFlex}, a cross-layer memory system that addresses
these challenges in a full-HBF organization. HBFlex couples base-die
SRAM buffering with co-located prefill and decode execution, allowing
prefill computation to provide opportunities for draining decode KV
updates. Its system design separates logical reuse granularity from
physical striping, coordinates KV placement with attention scheduling,
creates and exploits sufficiently long writeback windows, and combines
lifetime-aware block packing with deferred reclamation. We evaluate
HBFlex using a custom trace-driven simulator that models individual
read and write operations, across four LLMs~\cite{deepseekai2026deepseekv4,deepseekai2024deepseekv3,sun2024hunyuanlarge,yang2025qwen3}, four Alibaba Bailian
production traces~\cite{wang2025kvcachewild}, and SWE-bench workloads~\cite{jimenez2024swebench} with varying concurrency.
This paper makes four contributions:

\begin{itemize}
    \item We characterize the mismatches between dynamic KV management
    and HBF read, program, and erase operations, distinguishing
    placement imbalance from access imbalance and identifying the
    trade-offs among reuse, parallelism, write timing, and reclamation.

    \item We propose a full-HBF architecture with base-die SRAM and
    co-located prefill and decode, preserving HBF capacity and plane
    resources while buffering dynamic KV updates.

    \item We design three coordinated mechanisms: HBF-aware KV
    placement and scheduling, window-aware writeback, and lifetime-aware
    reclamation, enabling efficient KV management while retaining a
    fine-grained logical interface.

    \item Across the evaluated configurations, HBFlex achieves average throughput
    speedups of up to 1.58$\times$ over FlashAccel and 3.30$\times$ over H3, benefiting from higher HBF bandwidth and
    more efficient KV-cache management. Ablation
    studies quantify the benefits of read balancing, writeback
    scheduling, and lifetime-aware reclamation.
\end{itemize}

\section{Background}

\subsection{High-Bandwidth Flash Basics}

\begin{figure}
    \centering
    \includegraphics[width=\linewidth]{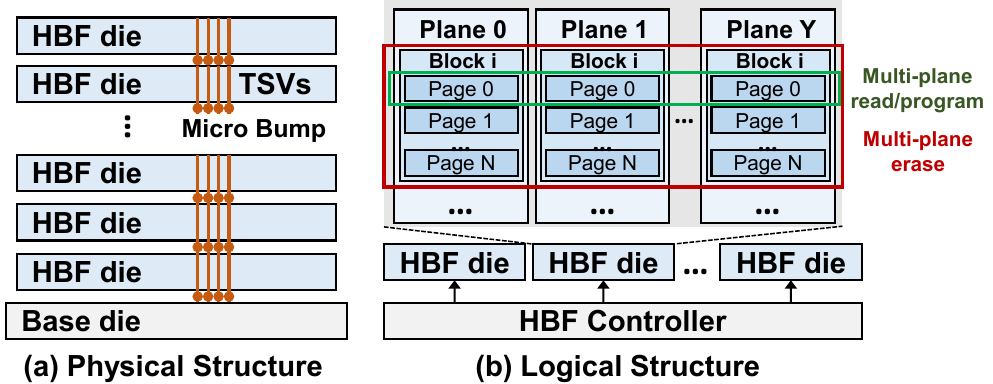}
    \caption{HBF organization: (a) stacked NAND dies and a base die;
    (b) the die--plane--block--page hierarchy.}
    \Description{A physical HBF stack with multiple NAND dies above a
    base die, and a logical hierarchy of dies, planes, blocks, and
    pages.}
    \label{fig:hbf_basic}
\end{figure}

High-Bandwidth Flash (HBF) combines NAND flash density and
non-volatility with a wide accelerator interface~\cite{ocp_hbf_architecture_v070}. As illustrated in
Figure~\ref{fig:hbf_basic}, NAND dies are stacked above a base die
that connects to the GPU and manages flash operations. HBF follows a
\emph{stack--die--plane--block--page} hierarchy, with each plane
providing an independent array datapath and page buffer. Array reads
and programs operate on 4-KiB physical pages, while erases operate on
blocks. The host interface may return the requested portion of
a buffered page.

HBF achieves high aggregate bandwidth through concurrent operations
across many planes. Striping pages across planes and host channels
exposes this parallelism, but sustaining peak bandwidth requires
sufficient outstanding requests and balanced access loads. Accesses
to the same plane serialize, so the most heavily loaded plane
determines completion time while less-loaded planes become idle.

\subsection{LLM Workload}
\label{sec:llm_workload}

LLM inference consists of \emph{prefill}, which processes prompt tokens
in parallel, and \emph{decode}, which generates tokens incrementally
while reusing cached KV vectors.

\begin{figure}
    \centering
    \includegraphics[width=\linewidth]{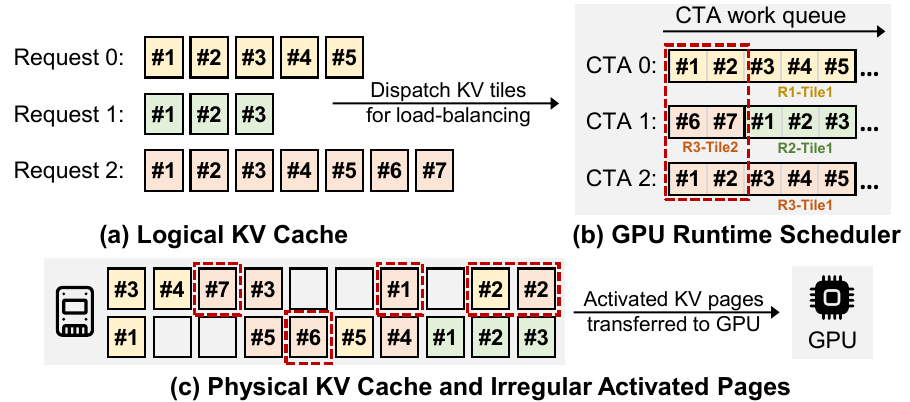}
    \caption{Paged KV-cache storage and chunked attention scheduling
    across CTAs.}
    \Description{Three requests with different KV-cache lengths are
    mapped to scattered physical pages and dispatched into a work queue
    processed by three CTAs.}
    \label{fig:attention}
\end{figure}

To support attention over this growing token history, PagedAttention
organizes KV state into small \emph{logical KV pages} mapped through a
page table to independently allocated physical pages~\cite{kwon2023pagedattention,
prabhu2025vattention}, as illustrated in
Figure~\ref{fig:attention}. A logically contiguous sequence therefore
spans fine-grained pages scattered across memory. At runtime, attention
kernels divide the sequence into \emph{work tiles}, or KV chunks, and
schedule them across cooperative thread arrays (CTAs). Each CTA follows
the page mappings to fetch its KV chunk, computes partial attention
results using the query vector, and contributes to a per-request
reduction~\cite{dao2024flashattention,ye2025flashinfer}. Consequently, a CTA's logically ordered work translates into
fine-grained reads from scattered, request-dependent physical locations,
rather than a single contiguous memory region.

\subsection{HBF for LLM Serving}

\begin{figure}
    \centering
    \includegraphics[width=\linewidth]{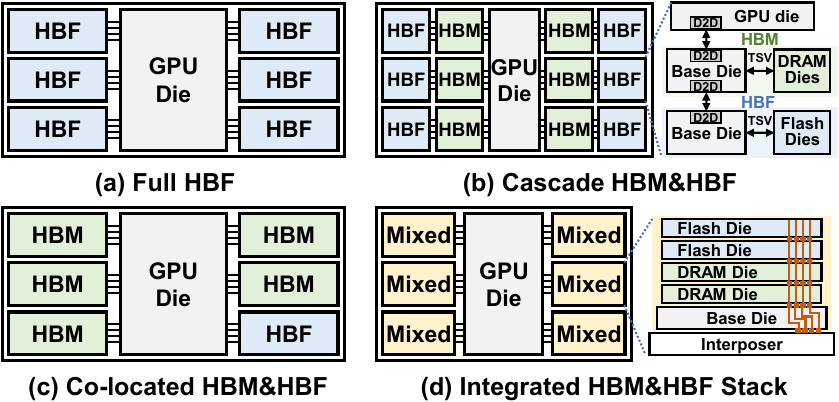}
    \caption{Representative HBF integration architectures.}
    \label{fig:hbf_arch}
\end{figure}

Recent architectures combine HBM and HBF to expand LLM serving
capacity, as illustrated in Figure~\ref{fig:hbf_arch}.
H3~\cite{haH3HybridArchitecture2026} connects HBF through the HBM base die, placing weights and
read-only shared KV in HBF while retaining generated KV in HBM.
FlashAccel~\cite{wangFlashAccelLeveragingHighBandwidth2026} explores both cascaded and co-located organizations,
storing weights and KV in HBF while using HBM for buffering and
read balancing. Integrated-stack designs~\cite{yin2026potentialapplicationshbfllm}
 instead combine DRAM and
flash dies within each stack, expanding capacity while preserving
HBM-like interfaces.

These designs retain HBM/DRAM to support dynamic or latency-sensitive
accesses. Under a fixed package or stack budget, this allocation
limits the resources available for HBF dies and parallel planes,
constraining both HBF capacity and aggregate bandwidth. Efficiently
managing dynamic KV directly in HBF could reduce this dependence on
HBM, but requires reconciling KV access and update patterns with
HBF's page-level operations and block-level reclamation.

\section{Motivation}

Efficiently accommodating dynamic KV cache in HBF requires aligning
KV management with the underlying flash organization. To understand
the challenges of such a design, we examine how fine-grained KV accesses
and dynamic updates interact with HBF's coarse-grained, plane-parallel
organization.

Our analysis uses an industry-provided HBF configuration with six
stacks, 16 dies per stack, and 32 planes per die, totaling 3,072 planes.
Each plane has 512 blocks of 512 pages each, with read and program
latencies of approximately 4 and 75~$\mu$s~\cite{wangFlashAccelLeveragingHighBandwidth2026}, respectively. The configured read bandwidth is 488~GB/s per stack, or 2.93~TB/s
across six stacks. Realizing
this bandwidth requires balanced accesses across planes. Dynamic KV
state complicates this requirement through changing active requests
and fine-grained reads, while incremental updates and mixed page
lifetimes introduce program interference and GC amplification. These
behaviors motivate the following three challenges in \textbf{KV read,
write, and erase behaviors}.

\begin{figure}[t]
    \centering
    \includegraphics[width=\linewidth]{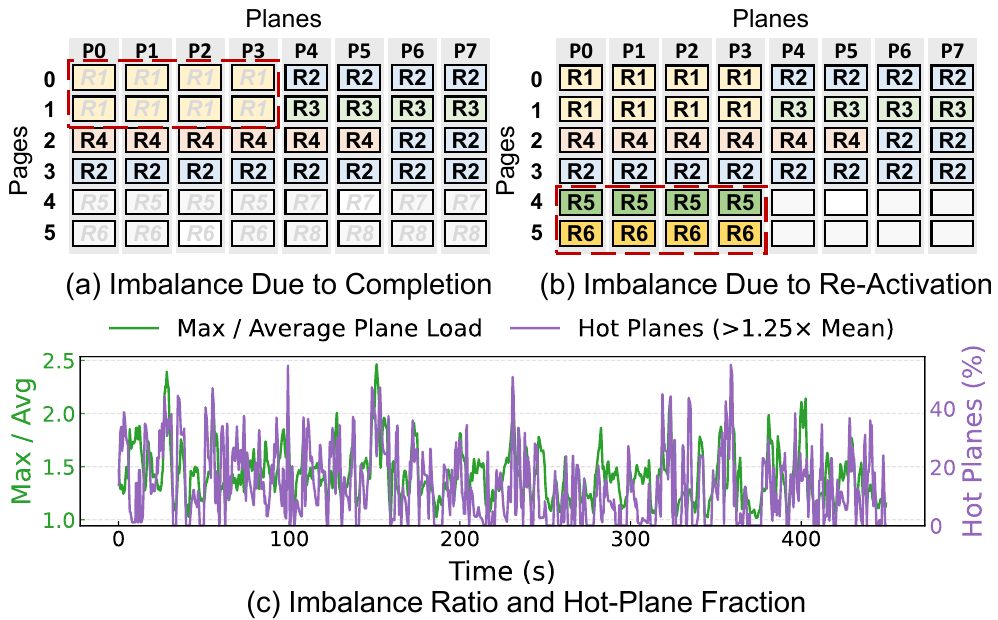}
    \caption{Placement imbalance from (a) request completion and
    (b) retained-KV reactivation. (c) Imbalance ratio and hot-plane
    fraction during a 450-s Bailian trace. Hot planes carry more than
    $1.25\times$ the mean load.}
    \Description{Two page maps illustrate placement imbalance caused
    by request completion and reactivation of retained KV pages. A time
    series below shows repeated spikes in the imbalance ratio and
    hot-plane fraction over a 450-second Bailian trace window.}
    \label{fig:dynamic_example}
\end{figure}

\subsection{Challenge 1: Fine-Grained KV Cache Creates Read Imbalance}

HBF read latency is determined by the most heavily loaded plane
because page reads within a plane serialize. We quantify imbalance
as the maximum plane load divided by the average:
\begin{equation}
    \rho = \frac{\max_i L_i}{\frac{1}{N}\sum_{i=1}^{N}L_i},
    \label{eq:read_imbalance}
\end{equation}
where $N$ is the number of planes and $L_i$ counts the relevant
physical pages on plane $i$. We measure this load over either the
active KV working set or the pages selected by an attention invocation.
These two scopes capture placement and access imbalance, respectively.
A ratio of one indicates perfect balance.

\paragraph{Placement imbalance.}
Request completion removes KV pages from the active working set without
necessarily deallocating them. This can expose imbalance among the
remaining requests. Allocating new pages to lightly loaded planes
repairs the current imbalance, but retained pages remain at their
original locations. Reactivating these pages can overload planes that
have since received new allocations, as illustrated in
Figure~\ref{fig:dynamic_example}(a) and (b).

We use traces from the Alibaba Bailian platform~\cite{wang2025kvcachewild} to examine placement
imbalance under our HBF configuration. As shown in
Figure~\ref{fig:dynamic_example}(c), the imbalance ratio and hot-plane
fraction exhibit repeated spikes throughout the trace. Balancing the current
active set therefore does not ensure balance after subsequent
completions and reactivations.

\begin{figure}[t]
    \centering
    \includegraphics[width=\linewidth]{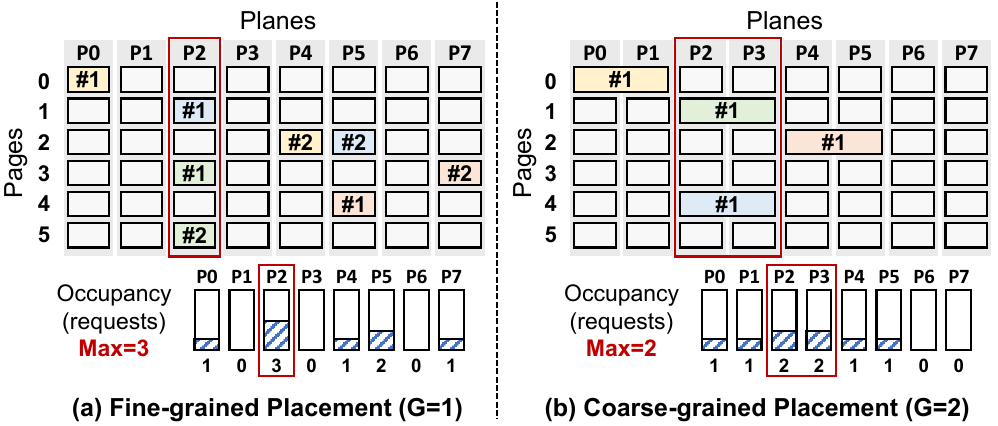}
    \caption{Access imbalance under balanced placement. Maximum plane
    occupancy decreases from three (a) to two (b) with two-page striping.
    $G$ denotes physical pages per stripe.}
    \Description{Two examples map pages from two requests onto eight
    planes. Independent page placement has a maximum plane load of
    three, whereas two-page striped placement has a maximum load of
    two.}
    \label{fig:occupancy}
\end{figure}

\paragraph{Access imbalance.}
Even with balanced placement, an attention invocation may access
planes unevenly. PagedAttention~\cite{kwon2023pagedattention} and chunked CTA scheduling~\cite{agrawal2023sarathi,
agrawal2024sarathiserve} select
fine-grained KV subsets whose pages can collide on the same plane.
For example, a 12-MiB read contains 3,072 physical pages, enough to
access every plane once under ideal placement. Collisions instead
force some planes to serve multiple pages sequentially. This requires
additional read waves, as illustrated in Figure~\ref{fig:occupancy}.

\begin{figure}[t]
    \centering
    \includegraphics[width=\linewidth]{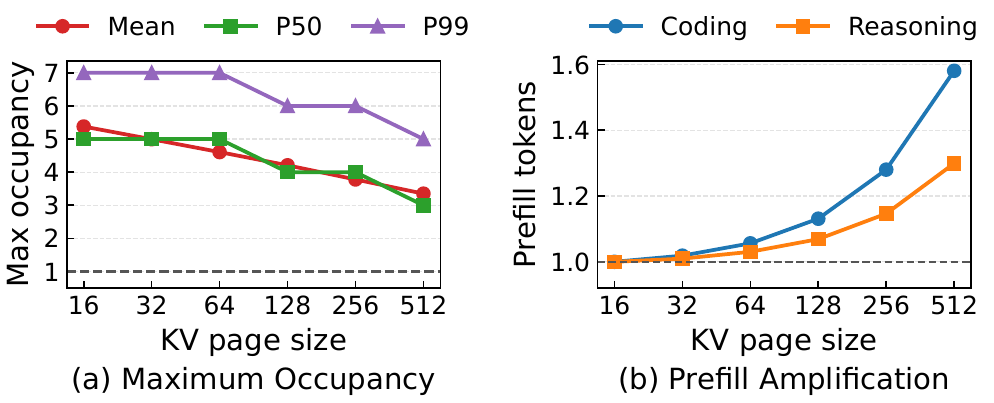}
    \caption{KV-page size trade-off under balanced placement. Larger
    pages reduce maximum plane occupancy for a 12-MiB read (a), but
    increase prefill amplification (b).}
    \Description{As KV page size increases from 16 to 512 tokens, the
    maximum plane occupancy decreases, but prefill-token amplification
    rises to 1.58 for coding and 1.30 for reasoning workloads.}
    \label{fig:occupancy_data}
\end{figure}

Increasing logical-page size can reduce these collisions, but it
coarsens reuse. Under the 1-KiB-per-token model, larger logical pages
span more physical pages striped across distinct planes. This reduces
independent placements and maximum plane occupancy in a 12-MiB read,
as shown in Figure~\ref{fig:occupancy_data}(a). However, prefix matching
reuses only complete logical pages. We define \emph{prefill amplification}
as the tokens recomputed with page-granular matching divided by the
minimum new tokens under token-granular matching. As shown in
Figure~\ref{fig:occupancy_data}(b), larger pages increase this
amplification by discarding reusable page tails. Tying logical reuse
granularity to physical access granularity therefore trades
fine-grained reuse for plane-level read parallelism.

\subsection{Challenge 2: Incremental KV Updates Create Write--Read Interference}
\label{sec:challenge-write}

Each generated token adds a small amount of KV state at every layer.
Writing these updates immediately underutilizes plane parallelism.
Buffering them enables wider writes, but increases buffer occupancy.
More importantly, each program occupies its target plane for
approximately 75~$\mu$s and cannot be preempted. Even an aggregated
write can therefore stall subsequent reads to the same planes.

\begin{figure}[t]
    \centering
    \includegraphics[width=\linewidth]{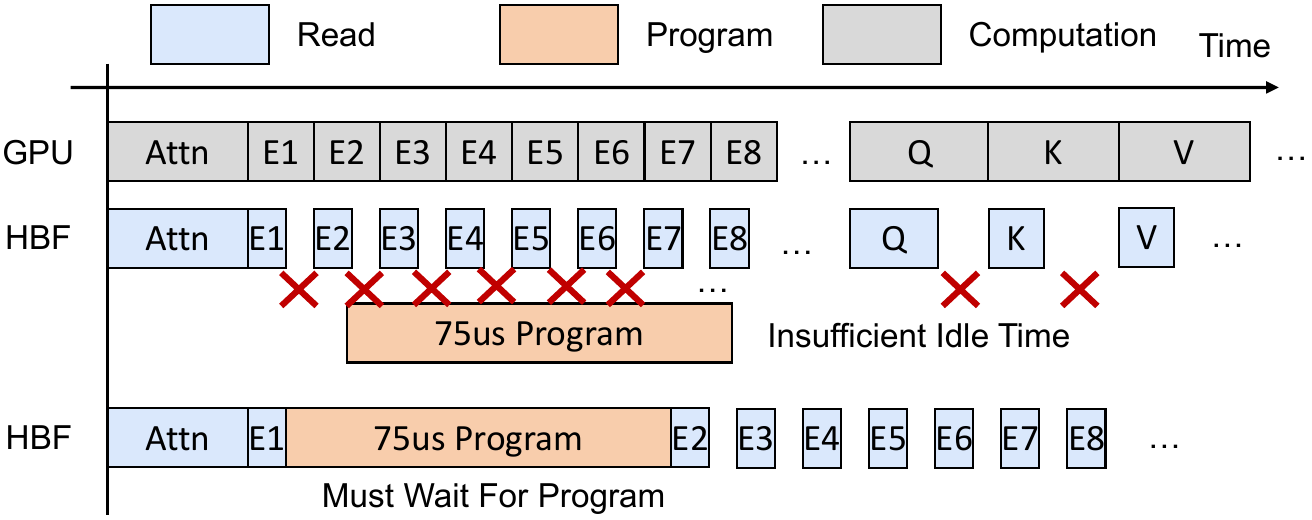}
    \caption{HBF program interference with foreground reads. Short
    compute gaps cannot hide a full program, causing subsequent reads
    to the same plane to stall.}
    \Description{A timeline contrasts short compute and HBF-read stages
    with a 75-microsecond program operation. Several candidate issue
    points are too short, while an issued program blocks later reads.}
    \label{fig:program}
\end{figure}

Hiding this latency requires a sufficiently long interval without
reads to the programming planes. However, weight and KV reads are
interleaved with computation, dividing potential write windows into
short gaps. As illustrated in Figure~\ref{fig:program}, a program that
extends beyond a gap blocks the next conflicting read. Deferring it
avoids the immediate stall, but buffer pressure eventually forces
writeback.

\begin{figure}[t]
    \centering
    \includegraphics[width=\linewidth]{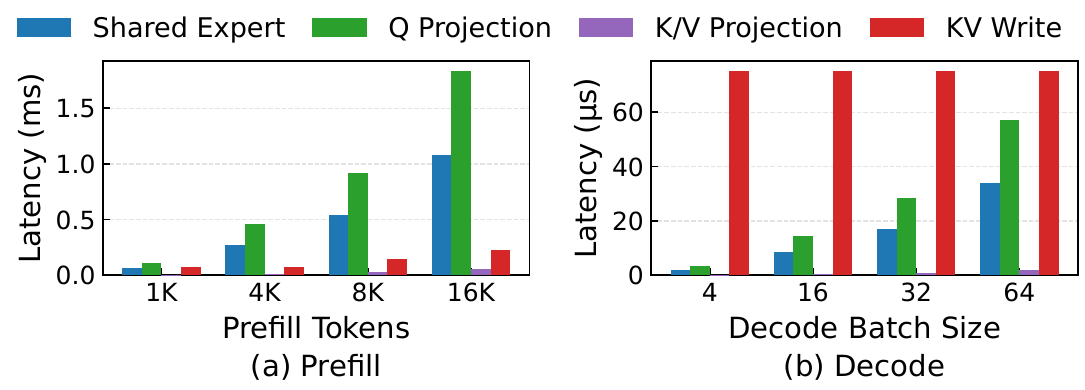}
    \caption{Computation and KV-write latency for DeepSeek-V4-Pro at an
    ideal 2,000 TOPS and a 75-$\mu$s HBF program latency. (a) Prefill
    with varying chunk sizes. (b) Decode with varying batch sizes.
    Intervening HBF reads are excluded.}
    \Description{Grouped bars compare shared-expert, Q-projection,
    K-or-V-projection, and KV-write latency across four prefill lengths
    and four decode batch sizes.}
    \label{fig:latency}
\end{figure}

As shown in Figure~\ref{fig:latency}, prefill computation grows with
chunk size, while most individual decode stages remain shorter than
one program. This comparison includes computation and KV writes but excludes
intervening HBF reads. Smaller writes occupy fewer planes but do not shorten the
program latency on each plane. Efficient writeback therefore requires
both aggregating updates and scheduling programs within sufficiently
long compute windows.

\subsection{Challenge 3: Interleaved KV Lifetimes Create Erase Amplification}
\label{sec:challenge-erase}

\begin{figure}
    \centering
    \includegraphics[width=\linewidth]{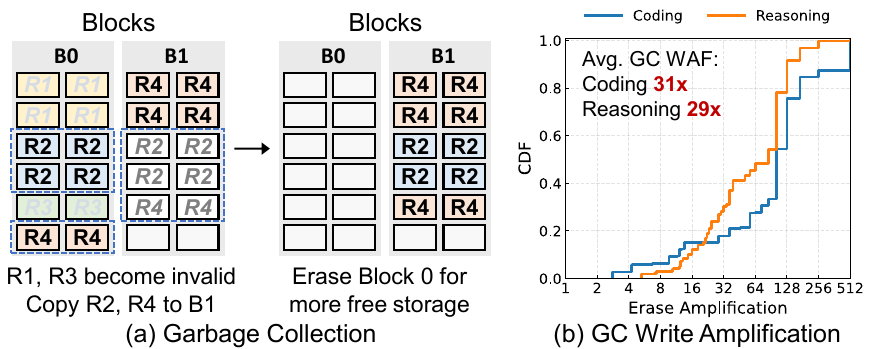}
    \caption{Garbage-collection amplification caused by interleaved KV lifetimes. (a) Valid KV pages must be migrated before block erase. (b) This effect leads to substantial GC amplification, with an average WAF of approximately 30$\times$ for DeepSeek-V4-Pro.}
    \label{fig:challenge_erase}
\end{figure}

\begin{figure*}[t]
    \centering
    \includegraphics[width=\linewidth]{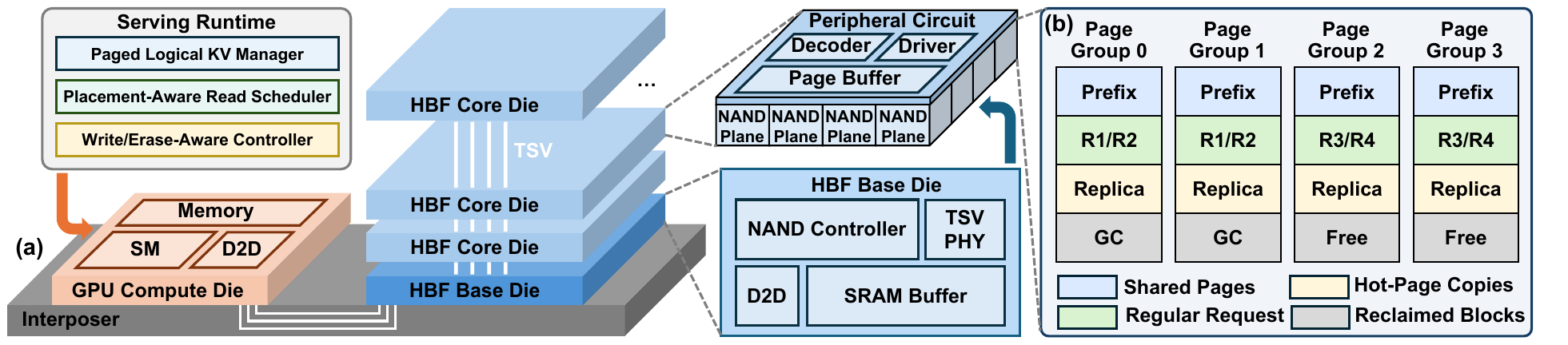}
    \caption{Overview of HBFlex architecture. (a) The GPU compute die connects to the HBF base die through D2D and the base die coordinates TSV-connected NAND core dies. (b) The schematic page-group layout with shared pages (blue), regular requests (green), hot-page copies (yellow), and reclaimed blocks (gray).}
    \label{fig:architecture}
\end{figure*}

Flash endurance is a potential concern when storing dynamic KV cache
in HBF~\cite{sonExploringHighBandwidthFlash2026}. DeepSeek reports approximately 434B new-KV tokens per day.
Using 35~KB/token for DeepSeek-V3~\cite{deepseekai2024deepseekv3}, this corresponds to 15.19~PB/day
system-wide, or 8.37~TB/day per GPU across 1,814 GPUs. At the same token
rate and GPU count, Qwen3~\cite{yang2025qwen3} would require 22.73~TB/day per GPU with
95~KB/token. A 1-TB HBF with 100K P/E cycles supports approximately
54.8~TB/day over five years. Raw KV writes consume 15.3\% of this budget
for DeepSeek-V3 and 41.5\% for Qwen3 under the same workload assumptions.
This leaves room for approximately 6.5$\times$ and 2.4$\times$ write
amplification, respectively.

The actual problem comes from HBF’s block-granular erase semantics~\cite{ocp_hbf_architecture_v070}. KV pages from different requests are interleaved in the same blocks but become invalid at different times. As shown in Figure~\ref{fig:challenge_erase}(a), when R1 and R3 expire, valid pages belonging to R2 and R4 must first be migrated before the block can be erased, introducing additional writes and erases through garbage collection. As shown in Figure~\ref{fig:challenge_erase}(b), the average GC WAF for DeepSeek-V4-Pro
reaches approximately 30$\times$. This exceeds the write-amplification
margins estimated above, making lifetime mixing a critical endurance
concern.

A seemingly simple solution is to place each request within dedicated blocks so that its KV cache can be reclaimed together. However, such request-isolated placement severely restricts plane-level striping and, as demonstrated in Challenge 1, creates substantial load imbalance and bandwidth loss. Therefore, KV-cache placement faces a fundamental tension: interleaving requests improves read parallelism, while isolating request lifetimes reduces erase amplification. Efficient HBF-based KV-cache management must jointly address both objectives.

\section{HBFlex Architecture}

In this work, we propose HBFlex, a cross-layer memory system that accommodates the dynamic KV cache of LLM serving in a full-HBF package while preserving the fine-grained logical interface that serving runtimes expose.
HBFlex co-locates prefill and decode on the same GPU--HBF package rather than using prefill--decode disaggregation, allowing prefill computation to provide HBF-free windows for draining decode KV writes.
By devoting the package's entire memory budget to HBF rather than retaining HBM, HBFlex maximizes capacity for the growing KV cache and preserves HBF's plane-level parallelism, supporting larger batch sizes and higher throughput without the HBF bandwidth loss incurred by a hybrid HBM/HBF package.
Section~\ref{sec:consider} develops the rationale for these choices and introduces the base-die SRAM buffer, while Section~\ref{sec:overview} describes the resulting architecture.

\begin{figure*}[ht]
    \centering
    \includegraphics[width=\linewidth]{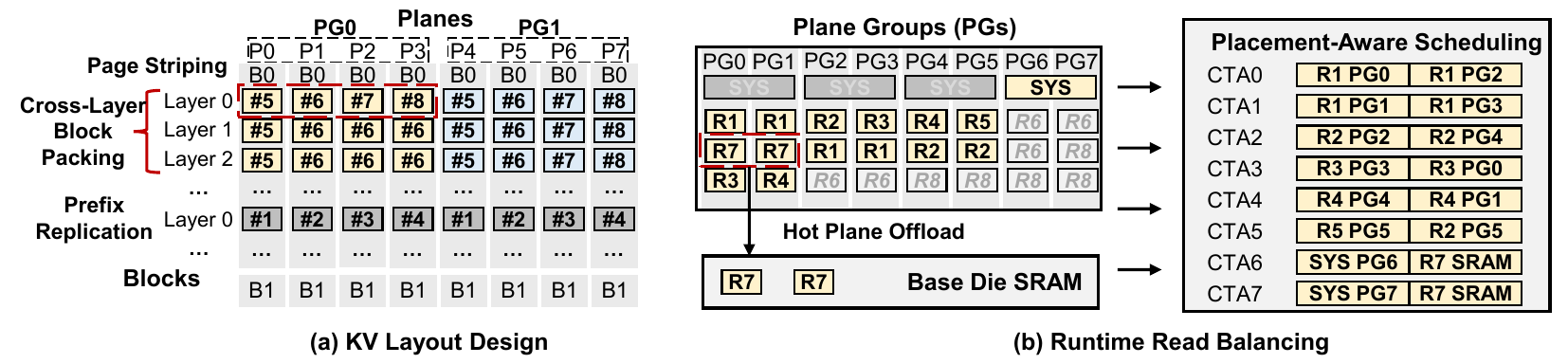}
    \caption{KV layout design and runtime read balancing in HBFlex. (a) Page striping, cross-layer block packing, and prefix replication organize KV cache across planes and blocks. (b) Runtime balancing mitigates hot-plane accesses through selective offloading and placement-aware scheduling.}
    \label{fig:system}
\end{figure*}

\subsection{Design Considerations}
\label{sec:consider}

\paragraph{Full-HBF package.}
Although flash endurance may appear to limit dynamic KV storage, the workload-level analysis in Section~\ref{sec:challenge-erase} shows that raw KV writes remain within the estimated endurance budget.
Garbage collection, rather than the raw write volume, is the dominant concern, with measured write amplification of approximately 30$\times$ for DeepSeek-V4-Pro.
HBFlex therefore dedicates the package to HBF rather than retaining HBM, which maximizes KV-cache capacity and preserves HBF's plane-level parallelism, supporting larger batch sizes and higher throughput.
The resulting placement, scheduling, write-formation, and reclamation challenges are addressed in Section~\ref{sec:sys}.

\paragraph{Fused prefill and decode.}
HBF reads are fast, massively parallel, and latency-critical, whereas programs are coarse, non-preemptible, and occupy a plane for about 75\,$\mu$s~(Section~\ref{sec:challenge-write}).
Because newly generated KV can be deferred, the scarce resource is a sufficiently long write window rather than aggregate write bandwidth.
Prefill--decode disaggregation separates continuous fine-grained decode writes from the compute-heavy prefill intervals that could absorb them.
HBFlex therefore co-locates both phases on one GPU--HBF package, buffers decode KV locally, and drains it during prefill when the target planes are off the critical path~(Section~\ref{sec:writeback}).

\paragraph{Base-die SRAM buffer.}
%Co-location makes buffering necessary, and the HBF base die can absorb updates across the D2D boundary without consuming GPU-visible memory or interconnect bandwidth.
The base-die SRAM is shared by HBF read prefetching, decode-KV writeback, hot-plane offload, and garbage-collection migration.
Its required capacity is determined by the peak simultaneous occupancy:
\begin{equation}
  C_{\mathrm{SRAM}}^{\mathrm{req}}
  = \max_t\!\left[
    C_{\mathrm{WPF}}(t)+C_{\mathrm{KV}}(t)+C_{\mathrm{aux}}(t)
  \right],
  \label{eq:buf}
\end{equation}
where $C_{\mathrm{WPF}}(t)$ is the live weight-prefetch footprint, $C_{\mathrm{KV}}(t)$ is the buffered decode-KV state, and $C_{\mathrm{aux}}(t)$ accounts for hot-plane offload and garbage-collection migration.
For DeepSeek-V4-Pro, most intervals between consecutive prefill opportunities accumulate less than 20\,MB of decode KV. This occupancy is added to the live weight-prefetch footprint when checking the shared SRAM capacity.
HBFlex provisions 40\,MB of SRAM per stack, or 240\,MB across six stacks, and forces KV writeback when the shared occupancy approaches the provisioned capacity.
At the 3\,nm technology node~\cite{wu20223nm}, a 40\,MB SRAM macro with 20\% peripheral overhead occupies approximately $8.06\,\mathrm{mm}^2$, or about 6.7\% of a $121\,\mathrm{mm}^2$ HBF base die~\cite{haH3HybridArchitecture2026}.
All evaluated designs use the same per-stack SRAM capacity so that comparisons reflect buffer management rather than SRAM provisioning.

\subsection{Architecture Overview}
\label{sec:overview}

Figure~\ref{fig:architecture}(a) shows the serving runtime and the three physical domains of HBFlex: the GPU compute die, the HBF base die, and the NAND core dies.
Prefill and decode execute on the GPU, which issues attention requests through D2D; the base die translates them through its NAND Controller and TSV PHY to selected core-die planes.
The base die also contains the SRAM Buffer, while each core die provides NAND planes and local peripheral circuits for selecting array locations and buffering page data.
The GPU consequently obtains aggregate plane-level HBF bandwidth without managing NAND commands, plane conflicts, or block reclamation.

Three runtime modules expose the logical control plane.
The Paged Logical KV Manager handles logical-page allocation, prefix reuse, request ownership, and validity.
The Placement-Aware Read Scheduler orders attention work tiles by expected plane load, and the Write/Erase-Aware Controller tracks dirty state and lifetime information for write and garbage-collection decisions.
HBFlex separates policy units through four explicit granularities, as illustrated by the page-group placement in Figure~\ref{fig:architecture}(b).
A \emph{logical KV page} is the runtime's allocation and prefix-reuse unit.
A \emph{physical stripe} groups $G$ adjacent physical pages across complementary planes without changing the logical page size.
A \emph{work tile} is the request-dependent KV subset consumed by one attention CTA.
An \emph{erase block} is the NAND reclamation unit whose valid pages must be copied before erase.
Mappings among these units allow reuse, placement, scheduling, and reclamation to be optimized independently.

% \subsection{Execution Flow}
% \label{sec:flow}
% The execution flow follows a cross-layer division of labor in which the serving runtime determines logical KV access and lifecycle actions, while the HBF base die executes plane-level reads, buffered programs, and block reclamation without exposing NAND management to the logical attention interface.
% For reads, the runtime schedules attention tiles across plane groups and redirects repeated accesses to hot-page copies or the SRAM Buffer when needed~(Section~\ref{sec:placement}).
% For writes, it aggregates incremental K/V state in the SRAM Buffer, forms wider stripes, and programs them during prefill windows or when dirty occupancy requires progress~(Section~\ref{sec:writeback}).
% For eviction, it uses lifecycle and lifetime information to select garbage-collection candidates, migrates valid pages through the SRAM Buffer, and erases the selected blocks~(Section~\ref{sec:packing}).

\section{HBFlex System Design}
\label{sec:sys}

\subsection{Design Overview}

To accommodate dynamic KV cache in HBF while preserving fine-grained
logical management, we propose three coordinated mechanisms that
address read imbalance, write--read interference, and erase
amplification. First, \emph{HBF-aware KV placement and scheduling}
coordinates physical page placement with runtime read balancing and
attention scheduling, improving plane utilization across both the
active KV working set and individual attention accesses. Second,
\emph{window-aware KV writeback} aggregates incremental updates in
base-die SRAM and schedules wider writes during sufficiently long
compute windows. By consolidating weight reads through prefetching and
exploiting prefill computation, it reduces interference between flash
programs and foreground reads. Third, \emph{lifetime-aware KV reclamation}
combines lifetime-guided block packing with deferred reclamation. It
groups pages with similar expected lifetimes within erase blocks and
allows more pages to become invalid before GC, reducing valid-page
migration and erase amplification.

\subsection{HBF-Aware KV Placement and Scheduling}
\label{sec:placement}

HBFlex combines physical KV layout with runtime read balancing to
address both placement and access imbalance. As illustrated in
Figure~\ref{fig:system}, the layout distributes KV pages across planes
and provides alternative locations for shared prefixes. The runtime
uses this layout to rebalance active KV accesses and schedule attention
chunks according to their physical placement.

\paragraph{Plane-Balanced Placement.}
HBFlex preserves small logical KV pages for fine-grained prefix reuse,
while organizing their backing physical pages to expose plane
parallelism. We partition HBF planes into \emph{Plane Groups (PGs)} and
stripe adjacent physical pages across distinct planes within each group.
This layout spreads accesses to neighboring pages across planes without
increasing the logical allocation or reuse granularity. HBFlex also
replicates frequently shared prefixes, such as system prompts, across
multiple PGs. These copies let the runtime serve the same logical pages
from different locations, providing flexibility to redirect reads as
the active working set changes.

\paragraph{Runtime Read Balancing.}
Request completion and reactivation can disrupt the balance of the
active KV working set. To address this placement imbalance, HBFlex
first redirects shared-prefix reads to replicas in less-loaded PGs,
without additional data movement. It then caches excess KV pages from
the remaining hot planes in base-die SRAM. By offloading only the excess
pages, HBFlex reduces load on these planes while limiting SRAM usage
and data movement.

Balancing the active working set does not ensure balanced accesses
within each attention invocation. HBFlex therefore makes chunk
construction and CTA scheduling placement-aware, addressing access
imbalance at runtime. For each scheduling wave, it greedily selects KV
chunks whose pages collectively cover as many distinct planes as
possible. Among candidates with similar plane balance, it favors a
chunk from the request previously processed by the CTA, enabling
query-vector reuse. This policy improves plane utilization while
preserving query locality and the original chunked-attention computation.

\subsection{Window-Aware KV Writeback}
\label{sec:writeback}

\begin{figure}
    \centering
    \includegraphics[width=\linewidth]{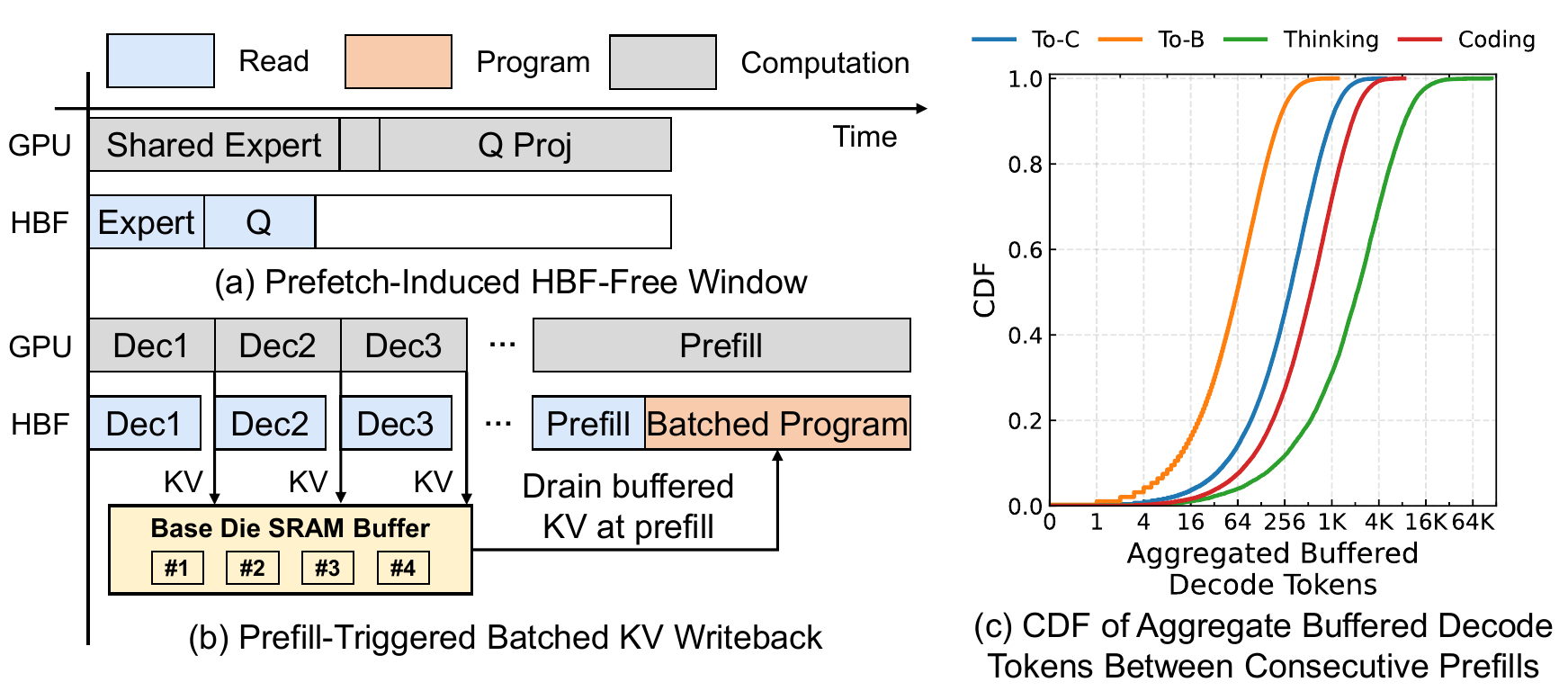}
    \caption{Window-aware KV writeback. (a) Prefetching creates an
    HBF-free compute window. (b) Batched decode-KV writeback during
    prefill. (c) CDF of decode tokens buffered between prefills across
    four Bailian traces.}
    \label{fig:method_write}
    %\vspace{-22pt}
\end{figure}

As discussed in Challenge~2, aggregating incremental KV updates
improves write parallelism, but does not by itself prevent interference
with foreground reads. HBFlex therefore decouples KV generation from
flash programming, buffering updates in base-die SRAM until a suitable
writeback opportunity arises. To make these opportunities available,
it combines proactive weight prefetching with the longer computation
offered by prefill.

Weight reads interleaved with linear operations fragment computation
into short intervals. To create longer writeback windows, HBFlex
prefetches the weights of consecutive operations into base-die SRAM,
as illustrated in Figure~\ref{fig:method_write}(a). These operations can
then run back-to-back without intervening weight reads, allowing
programming to overlap computation. It jointly
prefetches the three projections of an expert or the two Q-projection
matrices in models with low-rank Q projections. For DeepSeek-V4-Pro,
these groups occupy approximately 63 and 106~MB in INT8, respectively.
These weights share the 240-MB SRAM budget with buffered KV and other
temporary data, subject to the capacity constraint in
Section~\ref{sec:consider}.

Although prefetching consolidates computation, individual decode stages
may still be too short to hide a full program. HBFlex therefore retains
newly generated decode KV in SRAM and preferentially drains it during
prefill, as illustrated in Figure~\ref{fig:method_write}(b). Once the
required data have been fetched, the longer prefill computation provides
a suitable interval for writeback. This delay also allows updates from
multiple decode steps and requests to accumulate into wider programs,
improving plane-level write utilization.

Deferring decode writes until prefill requires sufficient SRAM to hold
the intervening updates. We assess this demand using the aggregate
number of decode tokens generated between consecutive prefill
opportunities. As shown in Figure~\ref{fig:method_write}(c), most
intervals accumulate fewer than 4K tokens across requests. At
approximately 5~KB/token for DeepSeek-V4-Pro, this corresponds to less
than 20~MB of buffered KV. When prefill opportunities
are sparse, shared occupancy may approach capacity before a suitable
window appears. HBFlex then forces writeback and temporarily stalls
decoding, ensuring bounded buffer usage.

\subsection{Lifetime-Aware KV Reclamation}
\label{sec:packing}

\begin{figure}
    \centering
    \includegraphics[width=\linewidth]{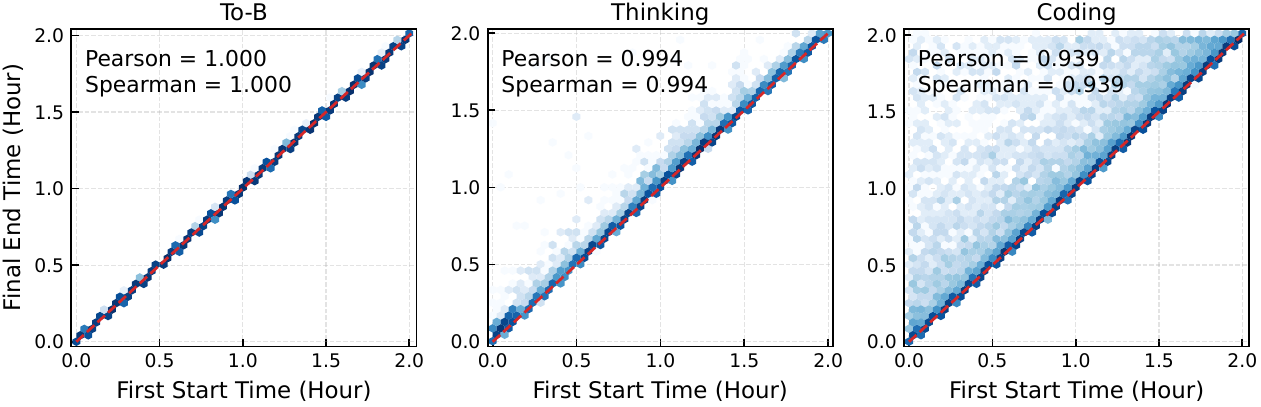}
    \caption{Correlation between request-series start and final end
    times across three workload traces. }
    \label{fig:method_lifetime}
    %\vspace{-15pt}
\end{figure}

To reduce the erase amplification identified in Challenge~3, HBFlex
coordinates lifetime-guided block packing with deferred reclamation.
Packing groups KV pages likely to become invalid together, while
deferred reclamation allows more invalid pages to accumulate before GC.
Both mechanisms reduce the valid data that must be migrated before
block erase.
KV pages from different Transformer layers of the same request have
closely related lifetimes, as they represent the same token history
and are typically evicted together. HBFlex exploits this relationship
by packing cross-layer KV pages into the same flash blocks whenever
possible, as illustrated in Figure~\ref{fig:system}(a).

Beyond this within-request correlation, HBFlex uses the root request's
start time to estimate lifetime similarity across request series. As
shown in Figure~\ref{fig:method_lifetime}, series that start at similar
times tend to have similar final end times. HBFlex therefore assigns
lifetime-interval labels based on root-request start times and propagates
them to descendant requests and their prefill/decode KV pages.
GC-relocated pages retain these labels to preserve lifetime grouping.

When allocating KV pages, HBFlex first looks for a partially filled
block with a matching lifetime-interval label. If none is available,
it opens a free block for that interval or uses a block with the
closest label. This policy groups pages with similar expected lifetimes
without requiring each request to occupy dedicated blocks, preserving
flexibility for plane-level striping.

Lifetime-guided packing improves the choice of pages within a block,
but reclaiming the block too early can still require unnecessary
migration. HBFlex therefore separates KV eviction from physical block
reclamation. It triggers eviction at 90\% memory utilization and GC
at 95\%, allowing additional pages to become invalid before blocks
are reclaimed. This interval gives GC more opportunities to select
blocks with fewer valid pages, further reducing migration overhead.

\begin{figure*}[!ht]
    \centering
    \includegraphics[width=\linewidth]{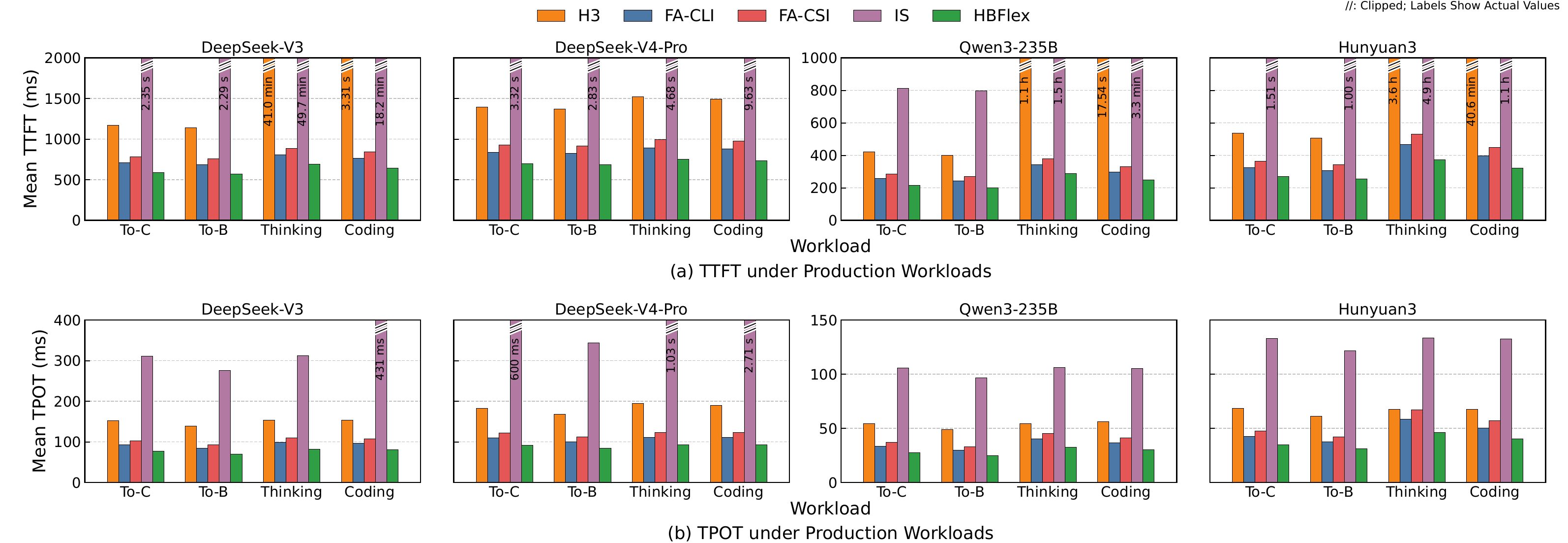}
    \caption{Serving latency across four models and four Alibaba Bailian traces: (a) mean TTFT and (b) mean TPOT. Labels above clipped bars indicate actual values.}
    \label{fig:performance}
\end{figure*}
\section{Experimental Evaluation}
\label{sec:eval}
\begin{figure*}
    \centering
    \includegraphics[width=\linewidth]{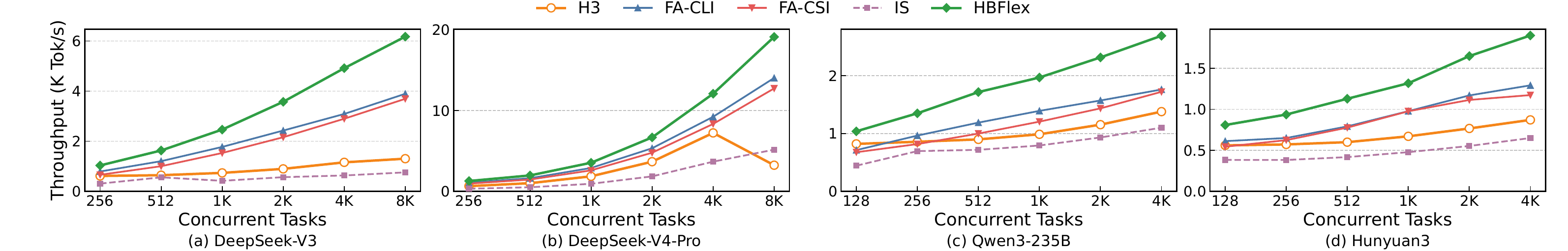}
    \caption{Serving throughput across four models as the number of concurrent tasks increases on SWE-bench traces.}
    \label{fig:throughput}
\end{figure*}
\subsection{Experimental Setup}

\textbf{Models and Workloads.}
We evaluate HBFlex on DeepSeek-V4-Pro~\cite{deepseekai2026deepseekv4},
DeepSeek-V3~\cite{deepseekai2024deepseekv3},
Hunyuan3~\cite{tencent2026hy3}, and
Qwen3-235B~\cite{yang2025qwen3}. DeepSeek-V4-Pro uses two-way tensor parallelism (TP) combined with
four-way data parallelism (DP), while the other models use four-way DP. Our workloads include four Alibaba
Bailian production traces~\cite{wang2025kvcachewild}: To-B, To-C, Coding, and Thinking. We also
collect SWE-bench~\cite{jimenez2024swebench} traces by running tasks with Qwen3 and recording the
inter-request intervals within each task. Varying task concurrency
allows us to evaluate performance under different serving loads.

\textbf{Hardware Configuration.}
We model H200-class accelerators~\cite{nvidia2023h200} with a common per-GPU compute
capability and memory-package area budget. The full-HBF configuration
contains six stacks per GPU. A full HBF stack contains 16 dies, with
32 independent planes per die. Each plane contains 512 blocks, and
each block contains 512 pages of 4\,KiB. The read and program latencies
are set to 4\,$\mu$s and 75\,$\mu$s, respectively. We configure each HBF stack with 488\,GB/s read bandwidth
and 27.2\,GB/s write bandwidth. Each HBF stack is equipped with a 40\,MB
base-die SRAM buffer for latency hiding and temporary data buffering.
We model an HBM3e stack with 8 DRAM dies, 24\,GB capacity, and
0.8\,TB/s bandwidth~\cite{jedec2023hbm3e} .

\textbf{Baselines.}
We compare HBFlex with representative HBF-based LLM serving designs,
including H3, FlashAccel, and an integrated-stack design (IS). H3
adopts a cascaded HBM/HBF organization, where read-mostly data are
placed in HBF while frequently updated states remain in HBM.
FlashAccel is evaluated under both co-located and cascaded
organizations. Following its original design, the co-located
configuration uses five HBF stacks and one HBM stack. We also evaluate
an integrated HBM/HBF stack organization, where HBM and HBF dies are
combined within the same stack.

For cascaded designs, we normalize the memory resources to the same
package-area budget. In the normalized configuration, one quarter of
the memory budget is assigned to HBM and three quarters to HBF.
Relative to full stacks, HBM capacity is scaled to one quarter, while
HBF capacity and read/write bandwidth are scaled to three quarters.
HBM bandwidth remains unchanged because the modeled external interface
width is preserved. HBFlex uses six full HBF stacks per GPU, with
base-die SRAM supporting dynamic KV management. All evaluated designs
use the same per-stack SRAM capacity.

\begin{table}[t]
\centering
\caption{Comparison of HBF-based LLM serving systems.}
\label{tab:baseline}
\resizebox{\columnwidth}{!}{
\begin{tabular}{cccc}
\toprule
\textbf{System} & \textbf{Architecture} & \textbf{KV Location} & \textbf{Key Feature} \\
\midrule
H3~\cite{haH3HybridArchitecture2026} & Cascaded & HBM+HBF & Prefix KV in HBF \\
FA-CLI~\cite{wangFlashAccelLeveragingHighBandwidth2026} & Co-located & HBM+HBF & Layout optimization \\
FA-CSI~\cite{wangFlashAccelLeveragingHighBandwidth2026} & Cascaded & HBM+HBF & Layout optimization \\
IS~\cite{yin2026potentialapplicationshbfllm} & Integrated Stack & HBM & Stack integration \\
HBFlex & Full HBF & HBF & Flexible Management \\
\bottomrule
\end{tabular}}
\end{table}

\subsection{End-to-End Performance}

We first evaluate serving latency using four Alibaba Bailian traces
across four models, as shown in Figure~\ref{fig:performance}. We report
geometric-mean speedups with equal weight for each model--trace
combination. HBFlex achieves time to first token (TTFT) speedups of 1.20$\times$ and
1.33$\times$ over FA-CLI and FA-CSI, respectively, with corresponding
time per output token (TPOT) speedups of 1.21$\times$ and 1.35$\times$. Compared with H3, the
average speedups reach 22.65$\times$ in TTFT and 1.89$\times$ in TPOT.
The larger TTFT gain reflects reduced admission queueing under
KV-capacity pressure, in addition to faster request processing.

To examine how these benefits vary with serving load, we use SWE-bench
traces with different numbers of concurrent tasks, as illustrated in
Figure~\ref{fig:throughput}. At lower concurrency, the smaller active KV
working set makes weight loading a larger component of memory service
time. Under the same package-area budget, HBFlex dedicates more
resources to HBF and exposes more planes for parallel weight reads.
At the lowest plotted concurrency, HBFlex achieves average throughput
speedups of 1.31$\times$ over FA-CLI and 1.47$\times$ over FA-CSI.
These gains primarily reflect faster weight loading, before KV capacity
and access efficiency become influential.

As concurrency increases, KV accesses account for a larger share of
memory traffic, making KV management increasingly important. Compared
with FlashAccel, which also stores KV in HBF, HBFlex gains additional
benefits from placement-aware attention scheduling and coordinated
writeback, improving read utilization while reducing write--read
interference. At the highest plotted concurrency, the average throughput
speedups over FA-CLI and FA-CSI increase to 1.48$\times$ and
1.58$\times$, respectively. Compared with H3, its larger HBF-resident KV capacity
allows more requests to remain active concurrently, relieving the
admission constraints imposed by HBM-resident KV. Across the four
models, the average throughput speedup over H3 increases from
1.56$\times$ at the lowest plotted concurrency to 3.30$\times$ at the
highest. These results illustrate how the benefits extend from faster
weight loading at low concurrency to more efficient KV access and
greater serving concurrency at higher loads.

\begin{table}[t]
    \centering
    \caption{Throughput speedups from serving simulation with
    cumulatively enabled HBFlex techniques, normalized to Naive HBF
    for each model.}
    \label{tab:cumulative_ablation}
    \resizebox{\columnwidth}{!}{%
    \begin{tabular}{lcc}
        \toprule
        \textbf{Cumulative configuration} & \textbf{DeepSeek-V3} & \textbf{Qwen3-235B} \\
        \midrule
        Naive HBF & 1.00$\times$ & 1.00$\times$ \\
        + Plane-Balanced Placement & 2.94$\times$ & 2.59$\times$ \\
        + Runtime Read Balancing & 4.08$\times$ & 4.25$\times$ \\
        + Window-Aware KV Writeback & 4.10$\times$ & 4.27$\times$ \\
        + Lifetime-Aware KV Reclamation & 4.15$\times$ & 4.32$\times$ \\
        \bottomrule
    \end{tabular}%
    }
\end{table}

\subsection{Ablation Study}
\label{sec:ablation}

We evaluate the cumulative throughput benefits of HBFlex through serving
simulation with 4,096 concurrent SWE-bench tasks across four
data-parallel GPUs. Starting from Naive HBF with immediate block-first
writes, we progressively enable each technique while retaining the
preceding optimizations.

As reported in Table~\ref{tab:cumulative_ablation}, most of the throughput
improvement comes from optimizing KV reads. Plane-balanced placement
and runtime read balancing together achieve speedups of 4.08$\times$
for DeepSeek-V3 and 4.25$\times$ for Qwen3-235B. Placement also improves
program parallelism, so this cumulative gain includes benefits to both
reads and writes.

Window-aware KV writeback provides a smaller additional gain, increasing
the normalized speedup by approximately 0.02$\times$. The relatively
small per-token KV footprints of the evaluated models limit write
traffic, leaving less write overhead to eliminate. Lifetime-aware KV
reclamation adds approximately 0.05$\times$, bringing the overall
speedups to 4.15$\times$ and 4.32$\times$. Its primary benefit is improved
flash endurance through fewer valid-page migrations and block erases,
while its contribution to throughput is more modest.

\subsection{Performance Analysis}
\label{sec:performance_analysis}

\paragraph{KV Placement and Scheduling.}

\begin{figure}
    \centering
    \includegraphics[width=\linewidth]{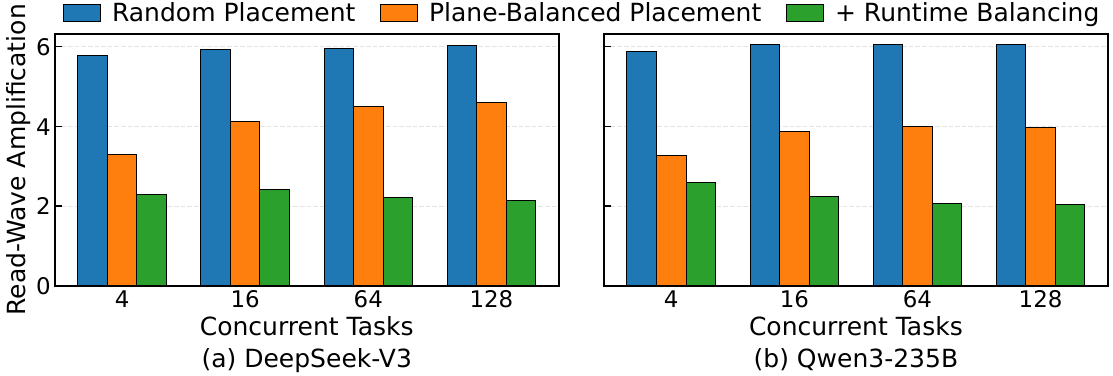}
    \caption{Effect of HBF-aware KV placement and runtime balancing on read-wave amplification.}
    \label{fig:ablation_read}
\end{figure}

We evaluate KV read efficiency using read-wave amplification, defined
as the ratio of actual read cycles to the theoretical minimum under
perfect plane utilization. As illustrated in Figure~\ref{fig:ablation_read},
we compare three configurations across four concurrency levels for
DeepSeek-V3 and Qwen3-235B. Averaged across all configurations of model
and concurrency, plane-balanced placement reduces amplification from
5.97$\times$ to 3.95$\times$, while runtime balancing further lowers it
to 2.25$\times$. Together, these mechanisms achieve an average reduction
of 62.21\% relative to random placement.

Balanced placement distributes stored KV pages across planes, but each
attention wave may still access only a subset of planes, leaving residual
amplification. Runtime balancing reduces these access conflicts through
selective offloading and placement-aware scheduling. The additional
improvement highlights the need to balance both stored pages and runtime
accesses.

\paragraph{Window-Aware KV Writeback.}

\begin{figure}[t]
    \centering
    \includegraphics[width=\linewidth]{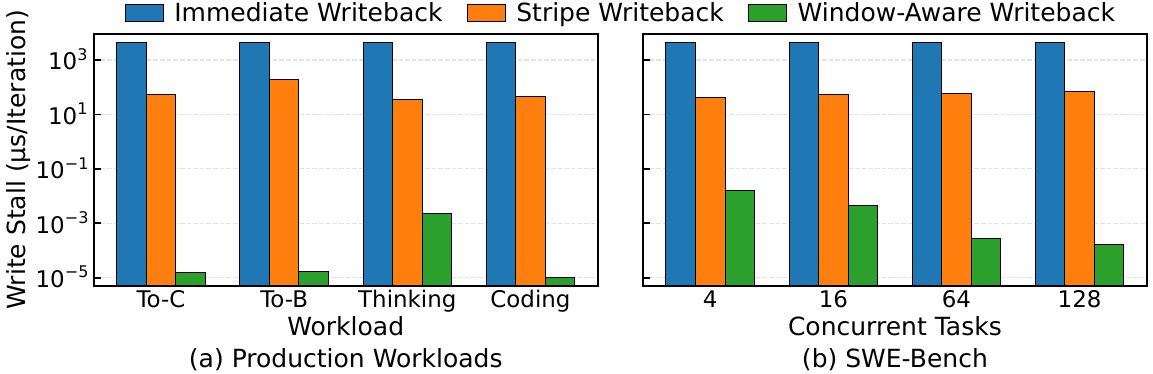}
    \caption{Effect of window-aware writeback on KV write stall.}
    \label{fig:ablation_write}
\end{figure}

We measure write-induced critical-path stall per serving iteration for
DeepSeek-V4-Pro across four Bailian traces and four SWE-bench concurrency
levels. As shown in Figure~\ref{fig:ablation_write}, the average stall
decreases from 4.54~ms with immediate writeback to 69.12~$\mu$s with
stripe writeback, and further to approximately 0.003~$\mu$s with
window-aware writeback.

Stripe writeback amortizes program overhead by combining small KV
updates into larger writes, but programs can still block subsequent
reads. Window-aware writeback adds weight prefetching and exploits
prefill compute windows to overlap programming with computation, nearly
eliminating stalls without shortening physical program latency. The
end-to-end gain remains modest when KV writes account for only a small
fraction of serving time.

\paragraph{Lifetime-Aware KV Reclamation.}

\begin{figure}[t]
    \centering
    \includegraphics[width=\linewidth]{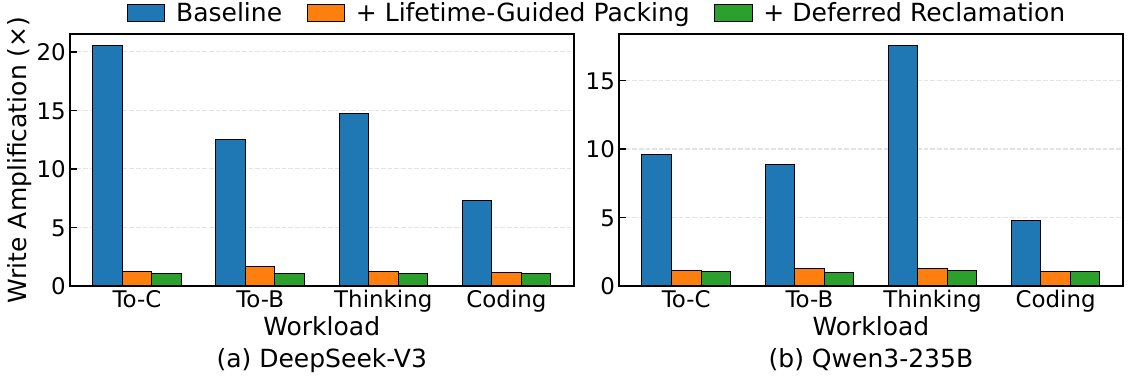}
    \caption{Effect of lifetime-guided packing and deferred reclamation on garbage-collection write amplification.}
    \label{fig:ablation_gc}
\end{figure}

We evaluate lifetime-aware KV reclamation using write amplification
(WAF)~\cite{hu2009write} across four Bailian traces for DeepSeek-V3 and
Qwen3-235B. As illustrated in Figure~\ref{fig:ablation_gc}, average WAF
decreases from 11.98$\times$ in the baseline to 1.24$\times$ with
lifetime-guided packing, and further to 1.04$\times$ with deferred
reclamation. The complete scheme achieves an average per-scenario
reduction of 89.39\% relative to the baseline.

Mixing KV lifetimes leaves valid pages in GC victims, requiring migration
before erase. Plane-group organization and lifetime-guided packing
concentrate invalidations within reclamation units, while deferred
reclamation lets more pages become invalid before GC. Together, they
reduce copying and bring WAF close to one. This limits extra writes and
associated erases, primarily benefiting endurance, alongside the modest
throughput gain observed in the cumulative ablation.

\subsection{Hot-Plane Offload and SRAM Usage}
\begin{figure}[t]
    \centering
    \includegraphics[width=\linewidth]{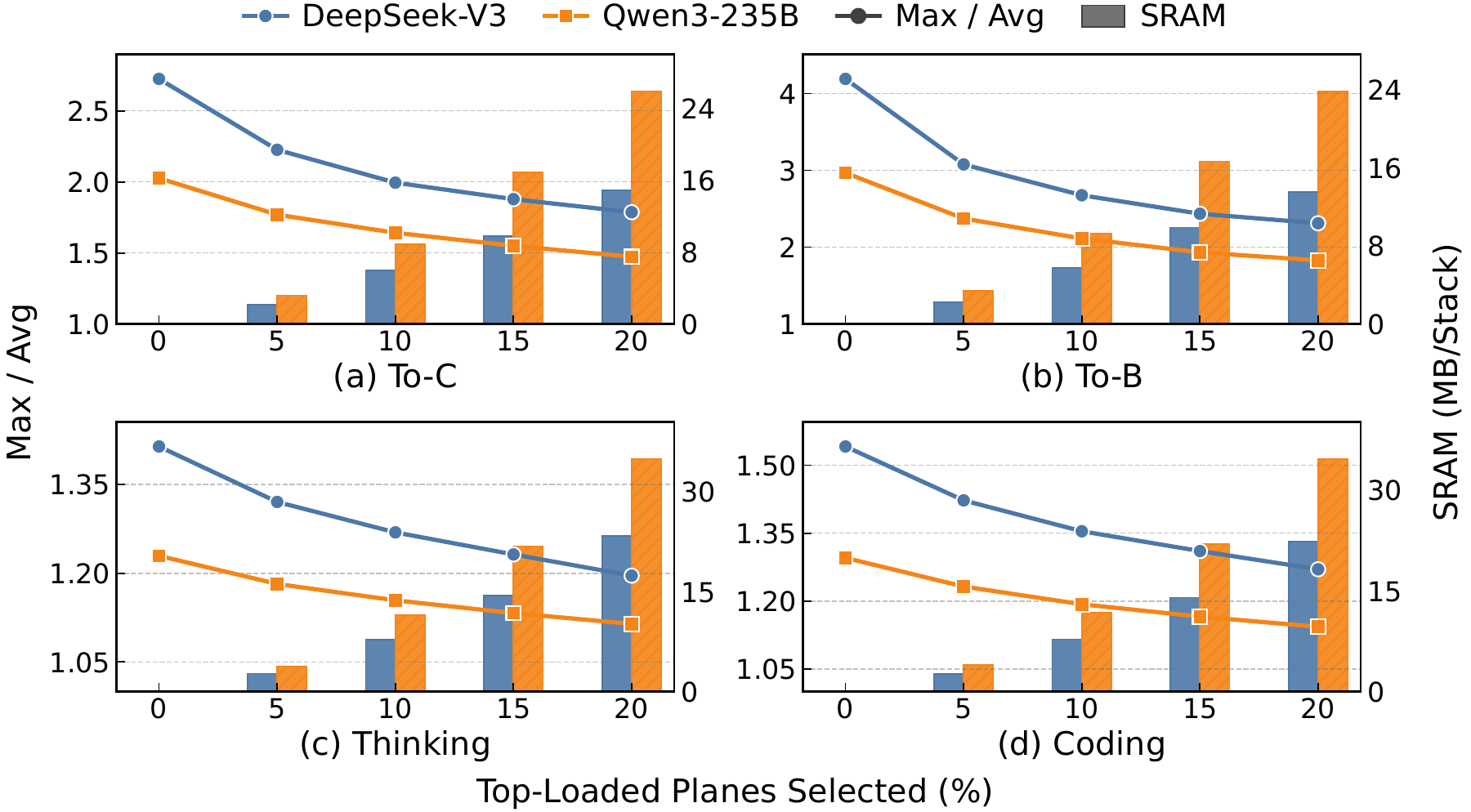}
    \caption{Trade-off between hot-plane offloading and SRAM usage
    across four Bailian traces. Lines show the maximum-to-average plane
    load; bars show SRAM usage per stack.}
    \Description{Four panels show plane load imbalance and SRAM usage
    for DeepSeek-V3 and Qwen3-235B on To-C, To-B, Thinking, and Coding
    traces. Selecting more top-loaded planes for offloading reduces
    imbalance but increases SRAM usage, with diminishing improvements
    in read balance beyond ten percent.}
    \label{fig:ablation_offload}
\end{figure}

We examine the trade-off between read balance and SRAM usage by varying
the fraction of top-loaded planes selected for offloading. The selected
fraction ranges from 0\% to 20\% under four Alibaba Bailian traces.
For each selected plane, excess KV pages are cached in base-die SRAM.
As illustrated in Figure~\ref{fig:ablation_offload}, we measure the
maximum-to-average plane load and the SRAM footprint per stack for
DeepSeek-V3 and Qwen3-235B.

Selecting more planes reduces load imbalance but increases SRAM usage.
The improvement is more pronounced on To-C and To-B, where the initial
imbalance is higher. Beyond 10\%, the curves flatten while SRAM usage
continues to grow. At 10\%, the offload footprint is approximately
6--12~MB per stack across the evaluated configurations. Increasing the
fraction to 20\% raises this footprint to approximately 14--35~MB per
stack. This leaves less SRAM for weight prefetching, KV writeback, and
GC migration. We therefore select the top-loaded 10\% of planes as a
trade-off between read balance and SRAM usage.

\section{Conclusion}
\label{sec:conclusion}

This paper presented HBFlex, a cross-layer memory system for serving
LLMs with dynamic KV cache in a full-HBF organization. Realizing HBF's
potential requires addressing placement and access imbalance,
write--read interference, and GC amplification from mixed KV lifetimes.
HBFlex combines
HBF-aware placement and attention scheduling, window-aware writeback,
and lifetime-aware reclamation to coordinate KV reads, writes, and
erases. Supported by base-die SRAM and co-located prefill and decode,
these mechanisms preserve fine-grained prefix reuse while adapting KV
management to HBF's physical constraints. Across the evaluated
configurations, HBFlex achieves average throughput speedups of up to
1.58$\times$ over FlashAccel and 3.30$\times$ over H3, benefiting from higher HBF bandwidth and more efficient
KV-cache management. These results demonstrate the benefits of
coordinating KV management to exploit full-HBF memory resources.

% use the ACM bibliography style
\bibliographystyle{ACM-Reference-Format}
\bibliography{sample-base}

@inproceedings{deshpande2025multichallenge,
  author    = {Deshpande, Kaustubh and Sirdeshmukh, Ved and Mols, Johannes Baptist and Jin, Lifeng and Hernandez-Cardona, Ed-Yeremai and Lee, Dean and Kritz, Jeremy and Primack, Willow E. and Yue, Summer and Xing, Chen},
  title     = {MultiChallenge: A Realistic Multi-Turn Conversation Evaluation Benchmark Challenging to Frontier LLMs},
  booktitle = {Findings of the Association for Computational Linguistics: ACL 2025},
  pages     = {18632--18702},
  year      = {2025}
}

@article{yi2025survey,
  author  = {Yi, Zihao and Ouyang, Jiarui and Xu, Zhe and Liu, Yuwen and Liao, Tianhao and Luo, Haohao and Shen, Ying},
  title   = {A Survey on Recent Advances in LLM-Based Multi-Turn Dialogue Systems},
  journal = {ACM Computing Surveys},
  volume  = {58},
  number  = {6},
  pages   = {1--38},
  year    = {2025}
}

@inproceedings{zheng2023judging,
  author    = {Zheng, Lianmin and Chiang, Wei-Lin and Sheng, Ying and Zhuang, Siyuan and Wu, Zhanghao and Zhuang, Yonghao and Lin, Zi and Li, Zhuohan and Li, Dacheng and Xing, Eric P. and Zhang, Haotong and Gonzalez, Joseph E. and Stoica, Ion},
  title     = {Judging LLM-as-a-Judge with MT-Bench and Chatbot Arena},
  booktitle = {Advances in Neural Information Processing Systems},
  volume    = {36},
  pages     = {46595--46623},
  year      = {2023}
}

@inproceedings{xiao2024efficient,
  title     = {Efficient Streaming Language Models with Attention Sinks},
  author    = {Xiao, Guangxuan and Tian, Yuandong and Chen, Beidi and Han, Song and Lewis, Mike},
  booktitle = {International Conference on Learning Representations (ICLR)},
  year      = {2024}
}

@inproceedings{zhang2023h2o,
  title     = {H2O: Heavy-Hitter Oracle for Efficient Generative Inference of Large Language Models},
  author    = {Zhang, Zichang and Sheng, Ying and Zhou, Tianyi and Chen, Tianlong and Zheng, Lianmin and Cai, Ruisi and Song, Zhao and Tian, Yuandong and R{\'e}, Christopher and Barrett, Clark and Wang, Zhangyang and Chen, Beidi},
  booktitle = {Advances in Neural Information Processing Systems (NeurIPS)},
  volume    = {36},
  pages     = {34661--34673},
  year      = {2023}
}

@inproceedings{wu20223nm,
  title = {A 3nm {CMOS} {FinFlex} Technology with Enhanced Power Efficiency and Performance for Mobile {SoC} and High Performance Computing Applications},
  author = {Wu, Shien-Yang and others},
  booktitle = {2022 IEEE International Electron Devices Meeting},
  pages = {27.1.1--27.1.4},
  publisher = {IEEE},
  address = {Piscataway, NJ, USA},
  year = {2022},
  url = {https://ieeexplore.ieee.org/document/10019483}
}

@inproceedings{park2023generative,
  author    = {Park, Joon Sung and O'Brien, Joseph and Cai, Carrie Jun and Morris, Meredith Ringel and Liang, Percy and Bernstein, Michael S.},
  title     = {Generative Agents: Interactive Simulacra of Human Behavior},
  booktitle = {Proceedings of the 36th Annual ACM Symposium on User Interface Software and Technology},
  pages     = {1--22},
  year      = {2023}
}

@article{wang2023voyager,
  author  = {Wang, Guanzhi and Xie, Yuqi and Jiang, Yunfan and Mandlekar, Ajay and Xiao, Chaowei and Zhu, Yuke and Fan, Linxi and Anandkumar, Anima},
  title   = {Voyager: An Open-Ended Embodied Agent with Large Language Models},
  journal = {arXiv preprint arXiv:2305.16291},
  year    = {2023}
}

@inproceedings{yao2023react,
  author    = {Yao, Shunyu and Zhao, Jeffrey and Yu, Dian and Du, Nan and Shafran, Izhak and Narasimhan, Karthik R. and Cao, Yuan},
  title     = {ReAct: Synergizing Reasoning and Acting in Language Models},
  booktitle = {The Eleventh International Conference on Learning Representations (ICLR)},
  year      = {2023}
}

@article{haH3HybridArchitecture2026,
  title = {{H3}: Hybrid Architecture Using High Bandwidth Memory and High Bandwidth Flash for Cost-Efficient {LLM} Inference},
  author = {Ha, Minho and Kim, Euiseok and Kim, Hoshik},
  journal = {IEEE Computer Architecture Letters},
  volume = {25},
  number = {1},
  pages = {49--52},
  year = {2026},
  doi = {10.1109/LCA.2026.3660969}
}

@article{kyungHighBandwidthFlashKV2026,
  title = {High-Bandwidth Flash for {KV} Caches: Endurance and Performance Implications},
  author = {Kyung, Kwanhee and Moon, Yeon Ji and Cho, Juhwan and Ahn, Jung Ho},
  journal = {IEEE Computer Architecture Letters},
  volume = {25},
  number = {1},
  pages = {210--213},
  year = {2026},
  doi = {10.1109/LCA.2026.3695938}
}

@inproceedings{leeMemVillageHybridHBFHBM2025,
  title = {{Mem-Village}: A Hybrid {HBF-HBM} Architecture on Glass for Ultra-Scale {AI} Inference Systems},
  author = {Lee, Hyuni and Kim, Haeyeon and Lee, JungHyun and Choi, Inyoung and Lee, Jaehun and An, Hyowon and Yang, Chaemin and Kim, Joungho},
  booktitle = {2025 IEEE Electrical Design of Advanced Packaging and Systems},
  pages = {1--3},
  publisher = {IEEE},
  address = {Piscataway, NJ, USA},
  year = {2025},
  doi = {10.1109/EDAPS66187.2025.11411753}
}

@misc{liHBFSucksFullStack2026,
  title = {{HBF Sucks}! A Full-Stack Characterization of High-Bandwidth Flash for {KV}-Centric {LLM} Serving},
  author = {Li, Zhuoran and Bian, Zhuohang and Huang, Xin and Zhao, Yibo and Sun, Guangyu and Zhuo, Youwei},
  year = {2026},
  eprint = {2608.11668},
  archivePrefix = {arXiv},
  primaryClass = {cs.AR},
  doi = {10.48550/arXiv.2608.11668},
  url = {https://arxiv.org/abs/2608.11668}
}

@article{parkHBMHBFCentricMemoryPooling2026,
  title = {{HBM-HBF}-Centric Memory Pooling Architecture with Custom Base Die for Terabyte-Scale {LLM} Inference},
  author = {Park, Junho and An, Hyowon and Suh, Haeseok and Yoon, Youngsu and Lee, Hyuni and Kim, Joungho},
  journal = {IEEE Computer Architecture Letters},
  volume = {25},
  number = {2},
  pages = {259--262},
  year = {2026},
  doi = {10.1109/LCA.2026.3703982}
}

@article{sonExploringHighBandwidthFlash2026,
  title = {Exploring High-Bandwidth Flash for Modern {LLM} Inference: Opportunities and Challenges},
  author = {Son, Dowon and Park, Yonggon and Cho, Hyunuk and Ham, Hyungkyu and Mutlu, Onur and Lee, Sungjin and Kim, Gwangsun and Park, Jisung},
  journal = {IEEE Computer Architecture Letters},
  volume = {25},
  number = {2},
  pages = {1--4},
  year = {2026},
  doi = {10.1109/LCA.2026.3705817}
}

@misc{wangFlashAccelLeveragingHighBandwidth2026,
  title = {{FlashAccel}: Leveraging High-Bandwidth Flash for High-Throughput {LLM} Inference},
  author = {Wang, Xinyu and Xue, Yalong and Sun, Xiaotian and Zhang, Xiaoyu and Dou, Chunmeng and Li, Xueqi and Chen, Xiaoming},
  year = {2026},
  eprint = {2607.10186},
  archivePrefix = {arXiv},
  primaryClass = {cs.AR},
  doi = {10.48550/arXiv.2607.10186},
  url = {https://arxiv.org/abs/2607.10186}
}

@misc{yin2026potentialapplicationshbfllm,
      title={Potential Applications of HBF in LLM Serving Systems},
      author={Yihan Yin and Yinlun Zhao and Zhixin Yun and Guanying Wu and Feng Zhu and Kai Tao and Shu Li and Fei Huang and Zhe Zhang and Shuangchen Li and Hongzhong Zheng},
      year={2026},
      eprint={2608.13127},
      archivePrefix={arXiv},
      primaryClass={cs.AR},
      url={https://arxiv.org/abs/2608.13127},
}

@inproceedings{prabhu2025vattention,
  title = {{vAttention}: Dynamic Memory Management for Serving {LLMs} without {PagedAttention}},
  author = {Prabhu, Ramya and Nayak, Ajay and Mohan, Jayashree and Ramjee, Ramachandran and Panwar, Ashish},
  booktitle = {Proceedings of the 30th ACM Symposium on Operating Systems Principles},
  pages = {294--308},
  year = {2024}
}

@inproceedings{kwon2023pagedattention,
  title = {Efficient Memory Management for Large Language Model Serving with {PagedAttention}},
  author = {Kwon, Woosuk and Li, Zhuohan and Zhuang, Siyuan and Sheng, Ying and Zheng, Lianmin and Yu, Cody Hao and Gonzalez, Joseph E. and Zhang, Hao and Stoica, Ion},
  booktitle = {Proceedings of the 29th Symposium on Operating Systems Principles},
  pages = {611--626},
  publisher = {Association for Computing Machinery},
  address = {New York, NY, USA},
  year = {2023},
  doi = {10.1145/3600006.3613165},
  url = {https://arxiv.org/abs/2309.06180}
}

@techreport{ocp_hbf_architecture_v070,
  author      = {{Open Compute Project}},
  title       = {OCP HBF Architecture Specification v0.7.0 FINAL},
  institution = {Open Compute Project},
  type        = {Technical Specification},
  number      = {v0.7.0},
  year        = {2023},
  url         = {https://www.opencompute.org/}
}

@techreport{jedec2023hbm3e,
  author      = {{JEDEC Solid State Technology Association}},
  title       = {High Bandwidth Memory (HBM3/HBM3e) DRAM Standard},
  institution = {JEDEC},
  type        = {JEDEC Standard},
  number      = {JESD238A},
  year        = {2023},
  url         = {https://www.jedec.org/standards-documents/docs/jesd238a}
}

@inproceedings{hu2009write,
  author    = {Hu, Xiao-Yu and Eleftheriou, Evangelos and Haas, Robert and Iliadis, Ilias and Poff, Beat},
  title     = {Write amplification analysis in flash-based solid state drives},
  booktitle = {Proceedings of the 2nd ACM International Systems and Storage Conference},
  series    = {SYSTOR '09},
  pages     = {1--9},
  year      = {2009},
  publisher = {ACM},
  address   = {New York, NY, USA},
  doi       = {10.1145/1534530.1534544}
}

@inproceedings{dirik2009performance,
  author    = {Dirik, Arijit and Jacob, Bruce},
  title     = {The Performance of {PC} Solid-State Disks ({SSDs})},
  booktitle = {Proceedings of the 36th Annual International Symposium on Computer Architecture (ISCA)},
  pages     = {278--289},
  year      = {2009},
  doi       = {10.1145/1555754.1555790}
}

@inproceedings{agrawal2023sarathi,
  title = {{SARATHI}: Efficient {LLM} Inference by Piggybacking Decodes with Chunked Prefills},
  author = {Agrawal, Amey and Panwar, Ashish and Mohan, Jayashree and Kwatra, Nipun and Gulavani, Bhargav S. and Ramjee, Ramachandran},
  booktitle = {Proceedings of the 29th ACM International Conference on Architectural Support for Programming Languages and Operating Systems, Volume 2},
  pages = {315--329},
  year = {2024}
}

@inproceedings{agrawal2024sarathiserve,
  title = {Taming Throughput-Latency Tradeoff in {LLM} Inference with {Sarathi-Serve}},
  author = {Agrawal, Amey and Kedia, Nitin and Panwar, Ashish and Mohan, Jayashree and Kwatra, Nipun and Gulavani, Bhargav S. and Tumanov, Alexey and Ramjee, Ramachandran},
  booktitle = {Proceedings of the 18th USENIX Symposium on Operating Systems Design and Implementation},
  pages = {111--126},
  year = {2024}
}

@inproceedings{aminabadi2022deepspeedinference,
  title = {{DeepSpeed Inference}: Enabling Efficient Inference of Transformer Models at Unprecedented Scale},
  author = {Aminabadi, Reza Yazdani and Rajbhandari, Samyam and Zhang, Minjia and Awan, Ammar Ahmad and Li, Cheng and Li, Du and Zheng, Elton and Rasley, Jeff and Smith, Shaden and Ruwase, Olatunji and He, Yuxiong},
  booktitle = {Proceedings of the International Conference for High Performance Computing, Networking, Storage and Analysis},
  pages = {1--15},
  year = {2022}
}

@misc{deepseekai2024deepseekv3,
  title = {{DeepSeek-V3} Technical Report},
  author = {{DeepSeek-AI}},
  year = {2025},
  eprint = {2412.19437},
  archivePrefix = {arXiv},
  primaryClass = {cs.CL},
  url = {https://arxiv.org/abs/2412.19437}
}

@misc{deepseekai2026deepseekv4,
  title = {{DeepSeek-V4}: Towards Highly Efficient Million-Token Context Intelligence},
  author = {{DeepSeek-AI}},
  year = {2026},
  eprint = {2606.19348},
  archivePrefix = {arXiv},
  primaryClass = {cs.CL},
  url = {https://arxiv.org/abs/2606.19348}
}

@inproceedings{jimenez2024swebench,
  title = {{SWE-bench}: Can Language Models Resolve Real-World {GitHub} Issues?},
  author = {Jimenez, Carlos E. and Yang, John and Wettig, Alexander and Yao, Shunyu and Pei, Kexin and Press, Ofir and Narasimhan, Karthik},
  booktitle = {International Conference on Learning Representations},
  year = {2024}
}

@inproceedings{li2023alpaserve,
  title = {{AlpaServe}: Statistical Multiplexing with Model Parallelism for Deep Learning Serving},
  author = {Li, Zhuohan and Zheng, Lianmin and Zhong, Yinmin and Liu, Vincent and Sheng, Ying and Jin, Xin and Huang, Yanping and Chen, Zhifeng and Zhang, Hao and Gonzalez, Joseph E. and Stoica, Ion},
  booktitle = {Proceedings of the 17th USENIX Symposium on Operating Systems Design and Implementation},
  pages = {663--679},
  publisher = {USENIX Association},
  address = {Boston, MA, USA},
  year = {2023},
  url = {https://arxiv.org/abs/2302.11665}
}

@article{maChallengesResearchDirections2026,
  title = {Challenges and Research Directions for Large Language Model Inference Hardware},
  author = {Ma, Xiaoyu and Patterson, David},
  journal = {Computer},
  volume = {59},
  number = {5},
  pages = {55--64},
  year = {2026},
  doi = {10.1109/MC.2026.3652916},
  url = {https://arxiv.org/abs/2601.05047}
}

@misc{nvidia2023h200,
  title = {{NVIDIA H200 Tensor Core GPU}},
  author = {{NVIDIA}},
  year = {2023},
  url = {https://www.nvidia.com/en-us/data-center/h200/},
  note = {Product specification, accessed 2026-09-08}
}

@inproceedings{patel2024splitwise,
  title = {{Splitwise}: Efficient Generative {LLM} Inference Using Phase Splitting},
  author = {Patel, Pratyush and Choukse, Esha and Zhang, Chaojie and Shah, Aashaka and Goiri, Inigo and Maleki, Saeed and Bianchini, Ricardo},
  booktitle = {Proceedings of the 51st Annual International Symposium on Computer Architecture},
  year = {2024}
}

@inproceedings{qin2025mooncake,
  title = {{Mooncake}: A {KVCache}-Centric Disaggregated Architecture for {LLM} Serving},
  author = {Qin, Ruoyu and Li, Zheming and He, Weiran and Zhang, Mingxing and Wu, Yongwei and Zheng, Weimin and Xu, Xinran},
  booktitle = {Proceedings of the 23rd USENIX Conference on File and Storage Technologies},
  year = {2025}
}

@misc{sun2024hunyuanlarge,
  title = {{Hunyuan-Large}: An Open-Source {MoE} Model with 52 Billion Activated Parameters by {Tencent}},
  author = {Sun, Yutao and others},
  year = {2024},
  eprint = {2411.02265},
  archivePrefix = {arXiv},
  primaryClass = {cs.CL},
  url = {https://arxiv.org/abs/2411.02265}
}

@misc{wang2025kvcachewild,
  title = {{KVCache Cache} in the Wild: Characterizing and Optimizing {KVCache Cache} at a Large Cloud Provider},
  author = {Wang, Jiahao and Han, Jinbo and Wei, Xingda and Shen, Sijie and Zhang, Dingyan and Fang, Chenguang and Chen, Rong and Yu, Wenyuan and Chen, Haibo},
  year = {2026},
  eprint = {2506.02634},
  archivePrefix = {arXiv},
  primaryClass = {cs.DC},
  url = {https://arxiv.org/abs/2506.02634}
}

@misc{yang2025qwen3,
  title = {{Qwen3} Technical Report},
  author = {Yang, An and others},
  year = {2025},
  eprint = {2505.09388},
  archivePrefix = {arXiv},
  primaryClass = {cs.CL},
  url = {https://arxiv.org/abs/2505.09388}
}

@inproceedings{zheng2024sglang,
  title = {{SGLang}: Efficient Execution of Structured Language Model Programs},
  author = {Zheng, Lianmin and Yin, Liangsheng and Xie, Zhiqiang and Sun, Chuyue and Huang, Jeff and Yu, Cody Hao and Cao, Shiyi and Kozyrakis, Christos and Stoica, Ion and Gonzalez, Joseph E. and Barrett, Clark and Sheng, Ying},
  booktitle = {Advances in Neural Information Processing Systems},
  volume = {37},
  pages = {62557--62583},
  publisher = {Curran Associates, Inc.},
  address = {Red Hook, NY, USA},
  year = {2024},
  url = {https://arxiv.org/abs/2312.07104}
}

@inproceedings{zhong2024distserve,
  title = {{DistServe}: Disaggregating Prefill and Decoding for Goodput-Optimized Large Language Model Serving},
  author = {Zhong, Yinmin and Liu, Shengyu and Chen, Junda and Hu, Jianbo and Zhu, Yibo and Liu, Xuanzhe and Jin, Xin and Zhang, Hao},
  booktitle = {Proceedings of the 18th USENIX Symposium on Operating Systems Design and Implementation},
  pages = {193--210},
  publisher = {USENIX Association},
  address = {Santa Clara, CA, USA},
  year = {2024},
  url = {https://arxiv.org/abs/2401.09670}
}

@article{ye2025flashinfer,
  title={Flashinfer: Efficient and customizable attention engine for llm inference serving},
  author={Ye, Zihao and Chen, Lequn and Lai, Ruihang and Lin, Wuwei and Zhang, Yineng and Wang, Stephanie and Chen, Tianqi and Kasikci, Baris and Grover, Vinod and Krishnamurthy, Arvind and others},
  journal={Proceedings of Machine Learning and Systems},
  volume={7},
  year={2025}
}

@inproceedings{dao2024flashattention,
  title={Flashattention-2: Faster attention with better parallelism and work partitioning},
  author={Dao, Tri},
  booktitle={International Conference on Learning Representations},
  volume={2024},
  pages={35549--35562},
  year={2024}
}

@misc{tencent2026hy3,
  author       = {{Tencent Hy Team}},
  title        = {Hy3},
  year         = {2026},
  howpublished = {\url{https://huggingface.co/tencent/Hy3}},
  note         = {Hugging Face model repository, accessed September 10, 2026}
}

\end{document}